\documentclass[twocolumn]{aastex61}
\usepackage{graphicx,subfigure,amsmath, amsfonts, amssymb,footnote,color,epstopdf}
\usepackage[bookmarks=false]{hyperref}
\usepackage{apjfonts}
\hypersetup{
  colorlinks= true, %Colours links instead of ugly boxes
  urlcolor  = black, %Colour for external hyperlinks
  linkcolor = red, %Colour of internal links
  citecolor = blue %Colour of citations
} % makes stupid green boxes go away, paper now looks more like it will when published!
\submitjournal{ApJS}
\accepted{4/11/2017}
\shorttitle{SERVS Multi-band Forced Photometry}
\shortauthors{Kristina Nyland et al.}
\begin{document}

%%%%%%%%%%%%%%%%%%%%%%%%%%%%%%%%%%%%%%%%%%%%%%%
% TITLE AND AUTHOR
\title{An Application of Multi-band Forced Photometry to One Square Degree of SERVS: Accurate Photometric Redshifts and Implications for Future Science}

\correspondingauthor{Kristina Nyland}
\email{knyland@nrao.edu}

\author{Kristina Nyland}
\affiliation{National Radio Astronomy Observatory, Charlottesville, VA 22903, USA}

\author{Mark Lacy}
\affiliation{National Radio Astronomy Observatory, Charlottesville, VA 22903, USA}

\author{Anna Sajina}
\affiliation{Department of Physics \& Astronomy, Tufts University, Medford, MA 02155, USA}

\author{Janine Pforr}
\affiliation{ESA/ESTEC SCI-S, Keplerlaan 1, 2201 AZ, Noordwijk, The Netherlands}
\affiliation{Aix Marseille Universit\'{e}, CNRS, LAM (Laboratoire d'Astrophysique de Marseille) UMR 7326, 13388, Marseille, France}

\author{Duncan Farrah}
\affiliation{Department of Physics, Virginia Tech, Blacksburg, VA 24061, USA}

\author{Gillian Wilson}
\affiliation{Department of Physics and Astronomy, University of California-Riverside, 900 University Avenue, Riverside, CA, 92521, USA}

\author{Jason Surace}
\affiliation{Spitzer Science Center, California Institute of Technology, M/S 314-6, Pasadena, CA 91125, USA}

\author{Boris H\"{a}u{\ss}ler}
%\affiliation{Oxford Astrophysics, Denys Wilkinson Building, Keble Road, Oxford OX1 3RH, UK; Centre for Astrophysics Research, University of Hertfordshire, College Lane, Hatfield AL10 9AB, UK}
\affiliation{European Southern Observatory, Alonso de Cordova 3107, Vitacura, Casilla 19001, Santiago, Chile}

\author{Mattia Vaccari}
%\affiliation{Astrophysics Group, University of the Western Cape, Private Bag X17, 7535, Bellville, Cape Town, South Africa}
\affiliation{Department of Physics and Astronomy, University of the Western Cape, Robert Sobukwe Road, 7535 Bellville, Cape Town, South Africa}
\affiliation{INAF - Istituto di Radioastronomia, via Gobetti 101, 40129 Bologna, Italy}

\author{Matt Jarvis}
\affiliation{Department of Physics and Astronomy, University of the Western Cape, Robert Sobukwe Road, 7535 Bellville, Cape Town, South Africa}
\affiliation{Department of Physics, Oxford Astrophysics, University of Oxford, Keble Road, Oxford OX1 3RH, UK}
%\affiliation{Physics Department, University of the Western Cape, Cape Town 7535, Republic of South Africa}

%Department of Physics, Oxford Astrophysics, University of Oxford, Keble Road, Oxford OX1 3RH, UK; Physics Department, University of the Western Cape, Cape Town 7535, Republic of South Africa

%%%%%%%%%%%%%%%%%%%%%%%%%%%%%%%%%%%%%%%%%%%%%%%
% ABSTRACT
\begin{abstract}
We apply {\it The Tractor} image modeling code to improve upon existing multi-band photometry for the {\it Spitzer} Extragalactic Representative Volume Survey (SERVS).  SERVS consists of post-cryogenic {\it Spitzer} observations at 3.6 and 4.5 $\mu$m over five well-studied deep fields spanning 18 deg$^2$.  In concert with data from ground-based near-infrared (NIR) and optical surveys, SERVS aims to provide a census of the properties of massive galaxies out to $z\approx5$.  To accomplish this, we are using {\it The Tractor} to perform ``forced photometry.''  This technique employs prior measurements of source positions and surface brightness profiles from a high-resolution fiducial band from the VISTA Deep Extragalactic Observations (VIDEO) survey to model and fit the fluxes at lower-resolution bands.  We discuss our implementation of {\it The Tractor} over a square degree test region within the XMM-LSS field with deep imaging in 12 NIR/optical bands.  Our new multi-band source catalogs offer a number of advantages over traditional position-matched catalogs, including 1) consistent source cross-identification between bands, 2) de-blending of sources that are clearly resolved in the fiducial band but blended in the lower-resolution SERVS data, 3) a higher source detection fraction in each band, 4) a larger number of candidate galaxies in the redshift range $5 < z < 6$, and 5) a statistically significant improvement in the photometric redshift accuracy as evidenced by the significant decrease in the fraction of outliers compared to spectroscopic redshifts.  Thus, forced photometry using {\it The Tractor} offers a means of improving the accuracy of multi-band extragalactic surveys designed for galaxy evolution studies.  We will extend our application of this technique to the full SERVS footprint in the future.
\end{abstract}

%%%%%%%%%%%%%%%%%%%%%%%%%%%%%%%%%%%%%%%%%%%%%%%
% SUBJECT HEADINGS

%%%%%%%%%%%%%%%%%%%%%%%%%%%%%%%%%%%%%%%%%%%%%%%
%%%%%%%%%%%%%%%%%%% INTRODUCTION %%%%%%%%%%%%%%%%%%%
%%%%%%%%%%%%%%%%%%%%%%%%%%%%%%%%%%%%%%%%%%%%%%%
\section{Introduction and Motivation}
Beginning with the successful SIRTF Wide-area Infrared Extragalactic Legacy Survey (SWIRE; \citealt{lonsdale+03}) over a decade ago, a growing number of deep extragalactic surveys with the {\it Spitzer Space Telescope} \citep{werner+04} have provided unprecedented insights into the evolution of galaxies over cosmic time.  Although the SWIRE footprint spanned $\sim$49 deg$^2$ and included imaging in seven {\it Spitzer} bands from the near- to far-infrared, it was primarily sensitive to objects at low and intermediate redshifts of $z \lesssim 3$ due to its relatively shallow depth.  More recently, the abundance of post-cryogenic observing time with {\it Spitzer} for large programs has made the construction of wide-area deep fields more feasible in the shortest two bands of the Infrared Array Camera (IRAC; \citealt{fazio+04}) that have remained in operation.  This has been particularly fortuitous for galaxy evolution studies since the combination of the inherent shapes of galaxy spectral energy distributions (SEDs) and the low background level in the IRAC bands make IRAC uniquely well-equipped for the detection of high-redshift galaxies given sufficient exposure time (e.g., \citealt{oesch+14}).  Furthermore, until the launch of the {\it James Webb Space Telescope} ({\it JWST}) currently scheduled for late 2018, IRAC is one of the few instruments capable of detecting rest-frame optical emission from galaxies at $z > 4$.  

%%%%%%%%%%%%%%%%%%%%%%%%%%%%%%%%%%%%%%%%%%%%%%%
\begin{figure*}
\centering
\includegraphics[clip=true, trim=0.1cm 0.25cm 0cm 0cm, width=14cm]{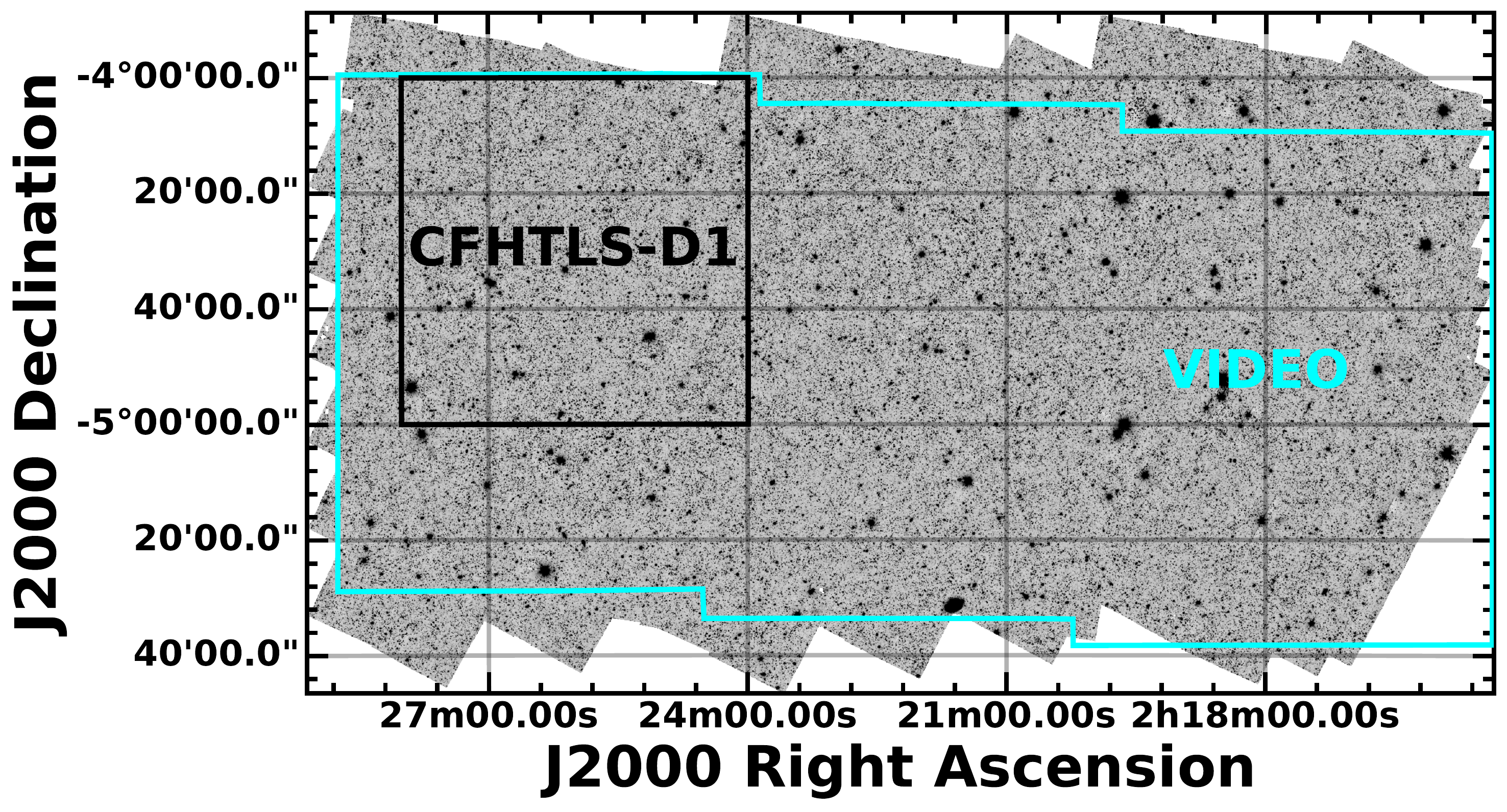}
\caption{{\it Spitzer} IRAC 3.6~$\mu$m image from SERVS \citep{mauduit+12} of the XMM-LSS field.  The cyan polygon traces the footprint of the VIDEO survey \citep{jarvis+13}.  The black square region denotes the location of the CFHTLS-D1 square degree footprint \citep{gwyn+12} over which we have tested the implementation of forced photometry with {\it The Tractor}.\\}
\label{fig:xmm_field}
\end{figure*}
%%%%%%%%%%%%%%%%%%%%%%%%%%%%%%%%%%%%%%%%%%%%%%

A number of recent surveys have capitalized on the opportunity to conduct deep, wide-area IRAC surveys at 3.6 and 4.5 $\mu$m during the warm phase of the {\it Spitzer} mission capable of detecting high-redshift galaxies.  Examples include the {\it Spitzer} IRAC Equatorial Survey (SpIES; \citealt{timlin+16}), the {\it Spitzer}-HETDEX Exploratory Large-Area Survey (SHELA; \citealt{papovich+16}) and the {\it Spitzer} Extragalactic Representative Volume Survey (SERVS; \citealt{mauduit+12}).  Here, we focus on the SERVS project, which 
%offers a combination of medium-depth and medium-area that 
is deep enough to detect $L_{*}$ galaxies at $z\approx5$, but still wide enough to detect interesting and rare objects such as quasars and ultraluminous infrared galaxies.  Thus, in concert with abundant ancillary data that include deep observations at optical, far-infrared, and radio wavelengths, SERVS will provide new insights into the cosmic star formation and supermassive black hole accretion histories of galaxies over the redshift range of $z \sim 0-5$.

The basis for the scientific success of surveys such as SERVS lies in the construction of robust multi-band source catalogs.  The existing SERVS source catalogs \citep{mauduit+12} were constructed using traditional photometric methods and software (e.g., SExtractor; \citealt{bertin+96}).  These methods typically employ an aperture photometry approach, in which fluxes are computed within a fixed elliptical aperture.  This technique generally yields single-band source catalogs of acceptable accuracy.  However, the suite of multi-wavelength SERVS data incorporate ancillary ground-based NIR and optical imaging at higher spatial resolution ($0\farcs8$) compared to that of the IRAC 3.6 and 4.5 $\mu$m bands ($1\farcs95$ and $2\farcs02$, respectively).  Mixed-resolution, multi-band catalogs are typically constructed by performing positional cross-matching between the individual source catalogs for each band within a pre-defined search radius (e.g., \citealt{vaccari+16}).  A drawback to this approach for SERVS is that sources that are clearly resolved in the higher-resolution, ground-based bands may appear ``blended'' together as a single source in the {\it Spitzer} IRAC imaging.  If not corrected, this will inevitably lead to incorrect source cross-identification between bands as well as less accurate flux measurements and photometric redshifts.  

In response to the need for more accurate multi-band photometry, a number of new tools have been developed in recent years.  These include software packages that use prior information from a band with high-resolution imaging to model the flux in lower-resolution bands such as T-PHOT (the successor to TFIT; \citealt{merlin+15, merlin+16}), PyGFIT \citep{mancone+13}, {\it XID+} \citep{hurley+17} and other applications of Bayesian cross-matching \citep{marquez+14, budavari+16}, {\it LAMDAR} \citep{wright+16}, and {\it The Tractor} \citep{2016ascl.soft04008L, lang+16}.  Each of these tools has different strengths and weaknesses in areas such as the available options for modeling sources (point-source/resolved surface brightness profile vs. elliptical aperture), fitting heuristics (maximum likelihood estimator vs.\ Bayesian inference), PSF characterization (single Gaussian/mixture of Gaussians vs. model image), algorithm speed, and accessibility to the user community.  

T-PHOT, PyGFIT, and {\it The Tractor} offer source surface brightness profile modeling capabilities, an important feature for producing accurate multi-band photometry of resolved sources in crowded, mixed-resolution datasets.  The most widely used software that includes analytic source modeling as an option is T-PHOT (e.g., \citealt{merlin+16b}).   However, T-PHOT has a number of stringent image formatting requirements, such as the need for perfectly aligned pixel boundaries between the low- and high-resolution images.  This complicates matters for users who wish to perform multi-band photometry using source catalogs and images from surveys with heterogeneous sky footprints.  
%PyGFIT also has some restrictions
%This is problematic for users who wish to perform multi-band photometry using source catalogs and images from surveys that do not include segmentation maps or other information needed to run T-PHOT in their public archives.  The situation for PyGFIT is similar.  
In addition, analytic surface brightness profile modeling in both T-PHOT and PyGFIT requires users to perform an additional step of producing a model image based on the highest-resolution band using software such as the {\sc GALFIT} \citep{peng+02}, thus increasing the computation time.  In contrast to T-PHOT and PyGFIT, {\it The Tractor} provides a greater degree of flexibility, simplicity, and customization opportunities (for details, see Section~\ref{sec:catalog} and \citealt{2016ascl.soft04008L}).  Thus, {\it The Tractor} is well-suited to the application of multi-band optical/NIR photometry considered in this study.

%T-PHOT, PyGFIT, and {\it The Tractor} offer source surface brightness profile modeling capabilities, an important feature for producing accurate multi-band photometry of resolved sources in crowded, mixed-resolution datasets.  The most widely used software that includes analytic source modeling as an option is T-PHOT (e.g., \citealt{merlin+16b}).  However, T-PHOT requires detailed inputs, such as segmentation maps, and also has a number of stringent image formatting requirements.  This is problematic for users who wish to perform multi-band photometry using source catalogs and images from surveys that do not include segmentation maps or other information needed to run T-PHOT in their public archives.  The situation for PyGFIT is similar.  In addition, analytic surface brightness profile modeling in both T-PHOT and PyGFIT requires users to perform an additional step of producing a model image based on the highest-resolution band using software such as the {\sc GALFIT} \citep{peng+02}, thus increasing the computation time.  In contrast to T-PHOT and PyGFIT, {\it The Tractor} provides a greater degree of flexibility, simplicity, and customization opportunities (for details, see Section~\ref{sec:catalog} and \citealt{2016ascl.soft04008L}).  Thus, {\it The Tractor} is well-suited to the application of multi-band optical/NIR photometry considered in this study.
%However, implementation of this capability requires users to perform an additional step of producing a model fits image using the {\sc GALFIT} two-dimensional source fitting software \citep{peng+02}

Here, we present new multi-band forced photometry incorporating data from 12 NIR and optical bands over a square degree of the XMM Large Scale Structure (XMM-LSS) field included in SERVS.  We describe the SERVS project in detail in Section~\ref{sec:SERVS} and the heuristics of our application of forced photometry using {\it The Tractor} in Section~\ref{sec:catalog}.  In Section~\ref{sec:results}, we compare the basic properties of our new multi-band forced photometric catalogs with the original input source catalogs constructed using traditional positional matching between bands.  We discuss the color and photometric redshift accuracy of our new forced photometry catalogs, and also consider prospects for future science applications, in Section~\ref{sec:discuss}.  We summarize our results in Section~\ref{sec:summary}.  Throughout this study we adopt a $\Lambda$CDM cosmology with $\Omega_{\mathrm{M}}$ = 0.3, $\Omega_{\Lambda}$  = 0.7, and H$_0$ = 70 km~s$^{-1}$ Mpc$^{-1}$.

\begin{deluxetable*}{cccCCC}[t!]
\tablecaption{Summary of Multi-band Data in the XMM-LSS Square Degree Test Field \label{tab:bands}}
\tablecolumns{6}
%\tablenum{1}
\tablewidth{0pt}
\tablehead{
\colhead{Band} & \colhead{Telescope} & \colhead{Survey} & \colhead{$\lambda$ ($\mu$m)} & \colhead{5$\sigma$ Threshold} & \colhead{Angular Resolution (arcsec)}\\
\colhead{(1)} & \colhead{(2)} & \colhead{(3)} & \colhead{(4)} & \colhead{(5)} & \colhead{(6)}
}
\startdata
Near Infrared & & & & & \\ 
\hline
$[4.5]$ & {\it Spitzer} & SERVS &  4.50 & 23.1 & 2.02\\ 
$[3.6]$ & {\it Spitzer} & SERVS &  3.60 & 23.1 & 1.95 \\ 
$K_{\mathrm{s}}$ & VISTA & VIDEO & 2.15 & 23.8 & 0.84\\ 
$H$  & VISTA & VIDEO & 1.65 & 24.1 & 0.88  \\ 
$J$ & VISTA & VIDEO & 1.25 & 24.4 & 0.87 \\ 
$Y$ & VISTA & VIDEO & 1.02 & 24.5 & 0.86 \\ 
$Z$ & VISTA & VIDEO & 0.88 & 25.7 & 0.88  \\ 
\hline
Optical & & &  & & \\ 
\hline
$z^{\prime}$ & CFHT & CFHTLS-D1 & 0.93 & 26.2 & 0.81 \\ 
$i^{\prime}$ & CFHT & CFHTLS-D1 & 0.78 & 27.4 & 0.76  \\ 
$r^{\prime}$ & CFHT & CFHTLS-D1 & 0.64 & 27.7 & 0.77  \\ 
$g^{\prime}$ & CFHT & CFHTLS-D1 & 0.47 & 27.9 & 0.83  \\ 
$u^{\prime}$ & CFHT & CFHTLS-D1 & 0.35 & 27.5 & 0.87 \\ 
\enddata
\tablecomments{Column 1: Observing band or filter name.  Column 2: Telescope name.  Column 3: Survey name.  Column 4: Central wavelength of observing band/filter.  Column 5: 5$\sigma$ source detection threshold.  Thresholds are taken from \citet{mauduit+12} for SERVS, \citet{jarvis+13} for VIDEO (2$^{\prime \prime}$ aperture), and \citet{gwyn+12} for CFHTLS-D1.  All measurements are in AB magnitudes.  Column 6: Typical angular resolution in the final survey images.  For the ground-based VIDEO and CFHTLS-D1 observations, the angular resolution refers to the typical seeing conditions.  For the SERVS data, the angular resolution refers to the FWHM of the post-cryogenic IRAC PRF from the IRAC Instrument Handbook.}
\end{deluxetable*}
%%%%%%%%%%%%%%%%%%%%%%%%%%%%%%%%%%%%%%%%%%%%%%%%

%%%%%%%%%%%%%%%%%%%%%%%%%%%%%%%%%%%%%%%%%%%%%%%%
%%%%%%%%%%%%%%%%%%%% SERVS %%%%%%%%%%%%%%%%%%%%%%%%
%%%%%%%%%%%%%%%%%%%%%%%%%%%%%%%%%%%%%%%%%%%%%%%%
\section{SERVS}
\label{sec:SERVS}
\subsection{Overview}
The SERVS sky footprint includes five well-studied astronomical deep fields with abundant multi-wavelength data spanning an area of $\approx18$ deg$^2$ and a co-moving volume of $\approx0.8$~Gpc$^3$.  The five deep fields included in SERVS are the XMM-LSS field, Lockman Hole (LH), ELAIS-N1 (EN1), ELAIS-S1 (ES1), and Chandra Deep Field South (CDFS).  SERVS provides NIR, post-cryogenic imaging in the 3.6 and 4.5 $\mu$m IRAC bands to a depth of $\approx2$~$\mu$Jy.  IRAC dual-band source catalogs generated using traditional catalog extraction methods are described in \citet{mauduit+12}.

The {\it Spitzer} IRAC data are complemented by ground-based NIR observations from the VISTA Deep Extragalactic Observations (VIDEO; \citealt{jarvis+13}) survey in the south in the $Z$, $Y$, $J$, $H$, and $K_{\mathrm{s}}$ bands and UKIRT Infrared Deep Sky Survey (UKIDSS; \citealt{lawrence+07}) in the north in the $J$ and $K$ bands.  SERVS also provides substantial overlap with infrared data from SWIRE \citep{lonsdale+03} and the {\it Herschel} Multitiered Extragalactic Survey (HerMES; \citealt{oliver+12}).  

Multi-band ``data fusion'' source catalogs for all five SERVS fields combining data from spectroscopic redshift surveys and photometry from the far-ultraviolet to the far-infrared using standard position-matching are described in \citet{vaccari+16}.  The SERVS data fusion multi-band catalogs are based on the IRAC 3.6 and 4.5~$\mu$m band-merged catalog presented in \citet{mauduit+12}.  For each SERVS source, the IRAC position is crossmatched with existing multi-wavelength source catalogs within a search radius of 1$^{\prime \prime}$.  For further details on the construction and contents of the SERVS data fusion catalogs, we refer readers to \citet{vaccari+16}.  

The suite of multiwavlength data available in the SERVS fields at NIR and optical wavelengths are especially well-suited for determining photometric redshifts for high-redshift objects (e.g., \citealt{ilbert+09}).  Photometric redshifts over the five SERVS fields based on the data fusion catalogs will be presented in Pforr et al.\ (in preparation).

%%%%%%%%%%%%%%%%%%%%%%%%%%%%%%%%%%%%%%%%%%%%%%%%
 \subsection{Square Degree Test Field}
As shown in Figure~\ref{fig:xmm_field}, one square degree of the XMM-LSS field overlaps with ground-based optical data from the Canada-France-Hawaii Telescope Legacy Survey Deep field 1 (CFHTLS-D1).  The CFHTLS-D1 region is centered at RA(J2000) = 02:25:59, Dec.(J2000) = $-$04:29:40, and includes imaging through the filter set $u^{\prime}$, $g^{\prime}$, $r^{\prime}$, $i^{\prime}$, and $z^{\prime}$.  Thus, in combination with the NIR data from SERVS and VIDEO that overlap with the CFHTLS-D1 region, multi-band imaging over a total of 12 bands is available (Table~\ref{tab:bands}).  Because of the abundant multi-band data in the CFHTLS-D1 field, we use this field to test our implementation of multi-band photometry with {\it The Tractor}.

%%%%%%%%%%%%%%%%%%%%%%%%%%%%%%%%%%%%%%%%%%%%%%%%
%%%%%%%%%%%%%%%%%%% CATALOG %%%%%%%%%%%%%%%%%%%%%%%
%%%%%%%%%%%%%%%%%%%%%%%%%%%%%%%%%%%%%%%%%%%%%%%%
\section{Method}
\label{sec:catalog}

%%%%%%%%%%%%%%%%%%%%%%%%%%%%%%%%%%%%%%%%%%%%%%%%
\subsection{\it The Tractor}
\label{sec:tractor}
{\it The Tractor} (\citealt{2016ascl.soft04008L, lang+16}) uses prior source positions, fluxes, and shape information from a high-resolution, ground-based NIR fiducial band to model and fit the flux in the remaining NIR and optical bands.  Fitting using {\it The Tractor} essentially optimizes the likelihood for the photometric properties of each source in each band given initial information on the source and image parameters.  Input image parameters for each band include a noise model, a point spread function (PSF) model, image astrometric information (WCS), and calibration information (e.g., the ``sky noise'' or rms of the image background).  The input source parameters include the source positions, brightnesses, and surface brightness profile shapes.  {\it The Tractor} proceeds by rendering the source model convolved with the image PSF model at each band and performs a least squares fit to the image data.  Since {\it The Tractor} is made freely available to the astronomical user community in the form of a {\sc Python} module, it does not have a front end user interface and users must write a driver script.  

%%%%%%%%%%%%%%%%%%%%%%%%%%%%%%%%%%%%%%%%%%%%%%%%
\subsection{Input Catalogs}
\label{sec:input_cats}
{\it The Tractor} must be supplied with an input catalog of prior positions.  To construct our input catalog, we first cross-matched the VIDEO\footnote{VIDEO source catalogs and images were obtained from the fourth data release available at http://horus.roe.ac.uk/vsa/.  All VIDEO data presented in this study are based on the VIDEO DR4.} and SERVS catalogs using TOPCAT \citep{taylor+05}, retaining all VIDEO sources as well as any SERVS source matches within a search radius of 1$^{\prime \prime}$ in our test field.  We then matched this VIDEO-SERVS catalog with the ground-based optical CFHTLS-D1 catalog, again retaining all VIDEO sources and any CFHTLS-D1 matches within a search radius of 1$^{\prime \prime}$.  The resulting VIDEO-selected, multi-band input catalog contains 12 NIR and optical bands (see Table~\ref{tab:bands}) and a total of 117,281 objects.  

\begin{figure}
\includegraphics[clip=true, trim=0.35cm 0cm 0cm 0cm, width=8.5cm]{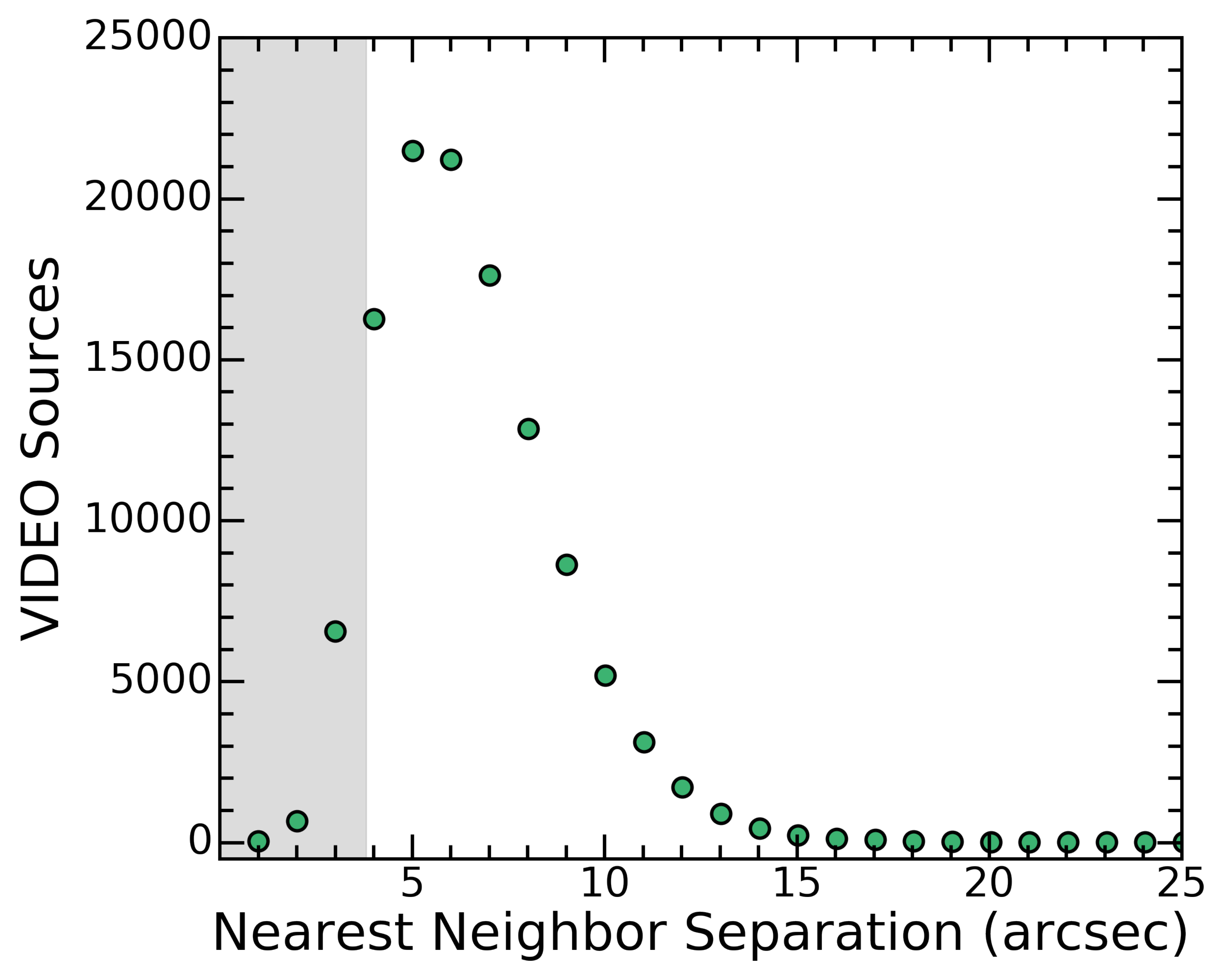}
\caption{Number distribution of nearest-neighbor VIDEO catalog source separations ($\Delta \theta$).  The grey-shaded portion highlights the population of VIDEO sources with $\Delta \theta < 3\farcs8$ that are expected to be blended in the original catalog of SERVS IRAC aperture photometry within a radius of $1\farcs9$, which corresponds to the FLUX\_APER\_2\_1 and FLUX\_APER\_2\_2 apertures for the 3.6 and 4.5 $\mu$m bands, respectively.}
\bigskip
\label{fig:VIDEO_blends}
\end{figure}

Based on the number distribution of nearest-neighbor source separations of $\Delta\theta \lesssim 3\farcs8$ (or about twice the angular resolution of the SERVS data) shown in Figure~\ref{fig:VIDEO_blends}, we expect at least 17\% of the 117,281 sources in the VIDEO-selected input catalog will be blended in the 3.6 and 4.5 $\mu$m IRAC data.  This is a lower limit since VIDEO sources with larger intrinsic angular sizes will be blended on even larger spatial scales in the original, position-matched SERVS photometric catalogs.   The high fraction of VIDEO sources expected to be blended in SERVS is one of the primary motivations for performing forced photometry with {\it The Tractor}.

To avoid biasing our output catalog against faint or extremely red objects that are detected only in the IRAC bands and have no ground-based NIR or optical counterparts, we also created a secondary input catalog of IRAC-selected sources\footnote{We emphasize that we perform forced photometry using {\it The Tractor} on both the VIDEO-selected and IRAC-selected input catalogs, thus producing two separate output multi-band source catalogs.}.  For this catalog, we included all detected SERVS sources lacking a counterpart within 1$^{\prime \prime}$ in the original VIDEO source catalog.  We also required a detection in at least one of the two IRAC bands in the SERVS single-band 3.6 and 4.5 $\mu$m catalogs\footnote{http://irsa.ipac.caltech.edu/data/SPITZER/SERVS/}.  The resulting IRAC-selected input catalog contains 8,441 sources.  

%%%%%%%%%%%%%%%%%%%%%%%%%%%%%%%%%%%%%%%%%%%%%%%%
\subsection{Fiducial Band Selection}
\label{sec:fiducial}
{\it The Tractor} generates a user-defined model of the surface brightness profile (see Section~\ref{sec:surf_bright_prof}) of a source at a given ``fiducial'' band with high-spatial-resolution imaging and then convolves this model with the PSF of each remaining, lower-resolution band.  Thus, the first step in our source modeling procedure is to determine the fiducial band.  For the VIDEO-selected catalog, we use the VIDEO $K_{\mathrm{s}}$-band data to define the fiducial high-resolution model of each source when possible since this band is closest in wavelength to the IRAC bands.  However, for sources in the VIDEO-selected catalog with non-detections at $K_{\mathrm{s}}$ band, we select a fiducial VIDEO band with a detection and valid flux entry with a preference for the band with the next closest central wavelength to the 3.6 $\mu$m IRAC data.  We note that, while we could select an optical band from the CFHTLS-D1 data as the fiducial band, this might result in the loss of very red objects from the catalog, and the possible misappropriation of infrared flux to unrelated galaxies with blue optical-infrared colors.

We follow a similar strategy for the IRAC-selected catalog of red sources detected in at least one IRAC band that lack a counterpart detected in any of the VIDEO bands in the original VIDEO catalog.  If a source is only detected in a single IRAC band, then that band is the fiducial band.  However, if both the 3.6 and 4.5 $\mu$m bands are detected, then we select the 3.6 $\mu$m band as the fiducial band.  The fiducial band selected for each source is provided in the Fiducial\_Band column (see Table~\ref{tab:cat_columns} in the Appendix) of both the VIDEO- and IRAC-selected output catalogs.

%%%%%%%%%%%%%%%%%%%%%%%%%%%%%%%%%%%%%%%%%%%%%%%%
%%%%%%%%%%%%%%%%%%%%%%%%%%%%%%%%%%%%%%%%%%%%%%%

%%%%%%%%%%%%%%%%%%%%%%%%%%%%%%%%%%%%%%%%%%%%%%%
\begin{deluxetable*}{ccCCCC}[t!]
\tablecaption{Summary of Image Calibration Parameters \label{tab:cal_params}}
\tablecolumns{6}
%\tablenum{2}
\tablewidth{0pt}
\tablehead{
\colhead{Band} & \colhead{Survey} & \colhead{Sky Noise} & \colhead{Sky Level} & \colhead{$\sigma_{\mathrm{Gaussian}}$} & \colhead{$w_{\mathrm{Gaussian}}$} \\
\colhead{(1)} & \colhead{(2)} & \colhead{(3)} & \colhead{(4)} & \colhead{(5)} & \colhead{(6)}
}
\startdata
Near Infrared & & & & & \\ 
\hline
$[4.5]$ & SERVS & 0.003 & 0.299 & [1.0, 1.95] & [0.3, 0.7] \\ 
$[3.6]$ & SERVS & 0.002 & 0.099 & [1.08, 2.20] &[0.37, 0.63] \\ 
$K_{\mathrm{s}}$ & VIDEO & 3.114 & -0.306 & [1.60, 3.09,10.0] & [0.59, 0.27, 0.135]\\ 
$H$  & VIDEO & 2.170 &  -0.226 & [1.60, 3.09, 8.64] & [0.61, 0.24, 0.15] \\ 
$J$ & VIDEO & 1.418 & -0.169 & [1.59, 2.94, 7.24] & [0.63, 0.30, 0.16] \\ 
$Y$ & VIDEO & 1.166 & -0.147 & [1.61, 3.15, 6.51] & [0.54, 0.19, 0.27] \\ 
$Z$ & VIDEO & 0.613 & -0.130 & [1.62, 2.76, 4.87, 11.80] & [0.5, 0.1, 0.32, 0.08] \\ 
\hline
Optical & & & & & \\ 
\hline
$z^{\prime}$ & CFHTLS-D1 & 0.774 & -0.117 & [1.56, 2.72, 6.21] & [0.73, 0.05, 0.23] \\ 
$i^{\prime}$ & CFHTLS-D1 & 0.330 & -0.096 & [1.27, 2.13, 4.35, 10.23] & [0.37, 0.42, 0.15, 0.06] \\ 
$r^{\prime}$ & CFHTLS-D1 & 0.248 &  -0.059 & [1.37, 2.27, 4.71, 11.45] & [0.36, 0.44, 0.16, 0.04] \\ 
$g^{\prime}$ & CFHTLS-D1 & 0.173 & -0.031 & [1.57, 2.71, 6.17] & [0.43, 0.44, 0.14] \\ 
$u^{\prime}$ & CFHTLS-D1 & 0.258 & -0.003 & [1.66, 2.84, 6.49] & [0.45, 0.42, 0.13] \\ 
\enddata
\tablecomments{Column 1: Observing band or filter name.  Column 2: Survey name.  Column 3:  Sky (rms) noise.  The values are given in native image units (counts for the VIDEO and CFHTLS-D1 images, and MJy~sr$^{-1}$ for SERVS).  Column 4:  Median background sky level.  Units are the same as in Column 3.  Column 5:  Standard deviation of each Gaussian component in our composite Gaussian models of the PSF of each band.  The mixture of Gaussians described by these models were used during source modeling and to estimate flux uncertainties.  We note that for SERVS we only used these  Gaussian PSF model parameters to estimate the flux uncertainties (the in-flight, post-cryogenic PSF model images were used instead during the source modeling stage).  Column 6:  The relative weights of the Gaussian components from Column 5, normalized to sum to 1.0.}
\end{deluxetable*}
%%%%%%%%%%%%%%%%%%%%%%%%%%%%%%%%%%%%%%%%%%%%%%%%

%%%%%%%%%%%%%%%%%%%%%%%%%%%%%%%%%%%%%%%%%%%%%%
\subsection{Surface Brightness Profile Modeling}
\label{sec:surf_bright_prof}
Once the fiducial band has been determined, we extract an image cutout of each source in the input VIDEO-selected catalog from the mosaicked image at each band using the {\sc Python} wrapper to the {\sc Montage}\footnote{http://montage.ipac.caltech.edu} toolkit \citep{berriman+17}, which is able to robustly interpret the complex WCS information in the headers of the {\it Spitzer} IRAC mosaics.  The resulting image cutouts each have a half-width of 5$^{\prime \prime}$.  This cutout size represents a trade-off between ensuring that the sources in our test field lie well within the cutout extent and excessive computational costs associated with larger cutout sizes.  Next, we create a fiducial band model of the target object as well as any neighboring sources in the VIDEO-selected input catalog that are present in the image cutout.

Based on the fiducial-band image, the source of interest along with neighboring sources within the cutout are modeled as either unresolved (i.e., a point source) or resolved.  For a source to be considered resolved, we require it to have a low probability of being a star in the VIDEO catalog (PSTAR $<$ 0.1) and an estimated radius $r > 0\farcs1$.  The radius is defined as $r_{\mathrm{source}} = \theta_{\mathrm{maj}} \times \sqrt{b/a} + 0.1$, where $\theta_{\mathrm{maj}}$ is the seeing-corrected half-light, semi-major axis (KSHLCORSMJRADAS for $K_{\mathrm{s}}$ band in the original VIDEO source catalog), $b/a$ is the axis ratio (semi-minor/semi-major) of an ellipse describing the source extent (determined from the KSELL VIDEO catalog parameter, where $b/a$ = 1 - KSELL), and the constant 0.1 refers to half the pixel size ($0\farcs2$) the VIDEO bands.

Photometry for resolved sources is then performed twice - once using a deVaucouleurs profile (equivalent to a Sersic profile with n = 4) and once using an exponential profile (equivalent to a Sersic profile with n = 1).  The resolved profile fit resulting in the lowest reduced chi squared ($\chi^{2}_{\mathrm{red}}$) value after optimization with {\it The Tractor} is reported in our final output catalog.

For the IRAC-selected catalog, we extract an image cutout from each band using {\sc Montage} and create a model of the source and its neighbors.  However, since the sources in the IRAC-selected catalog are typically near the SERVS detection limit, we restrict the source surface brightness profile models to be unresolved point-sources.

%%%%%%%%%%%%%%%%%%%%%%%%%%%%%%%%%%%%%%%%%%%%%%%%
\subsection{Image Calibration Parameters}
After the source model at the fiducial band has been determined, this model is convolved with the appropriate PSF for each band/instrument.  We use a mixture of circular Gaussians with 2-4 components each to model the PSFs for the ground-based VIDEO and CFHTLS-D1 data.  For each VIDEO and CFHTLS-D1 band, we select sources that are likely to be stars based on their bright (but unsaturated) fluxes, Gaussian-like radial profiles, and high PSTAR values.  We estimate the number of composite Gaussians needed to describe the source as well as the Gaussian $\sigma$ values and relative weights by visual inspection.  Finally, we use these estimates as initial guesses to obtain least-squares fits to the multi-component Gaussian PSF parameters for each band using {\it The Tractor}.  These parameters are listed in Table~\ref{tab:cal_params}.  For the SERVS data, the large wings and strong diffraction spikes of the PSF from these diffraction-limited, space-based data led to us using the in-flight post-cryogenic IRAC point response functions (PRFs) described in \citet{hora+12}. 

We also specify the sky (rms) noise and the median background sky level for each band.  These parameters correct for image contamination from sources such as instrument noise and zodiacal light.  The sky noise and sky level values used in each band are listed in Table~\ref{tab:cal_params}.

%%%%%%%%%%%%%%%%%%%%%%%%%%%%%%%%%%%%%%%%%%%%%%%
\begin{figure}
\centering
\includegraphics[clip=true, trim=0cm 0cm 0cm 0.25cm, height=20cm]{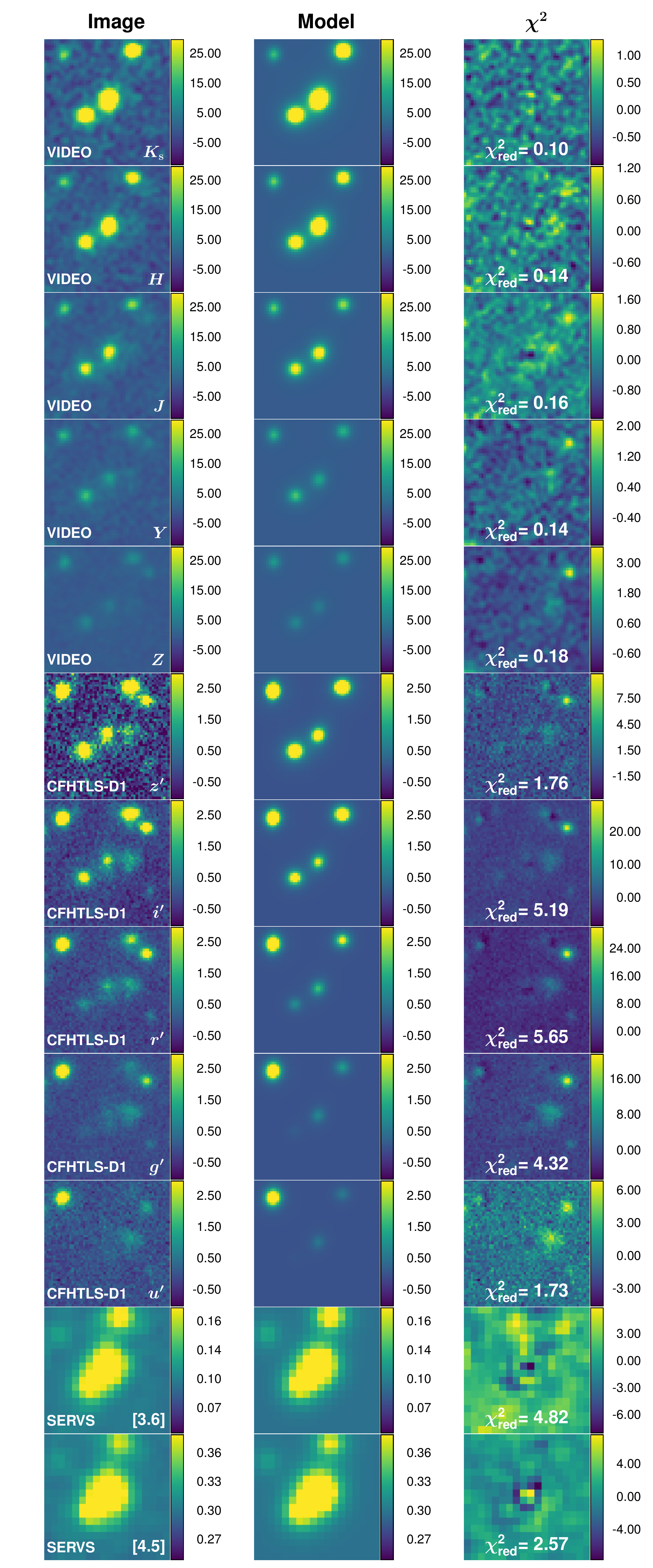}
\caption{Example of a source that is clearly blended in the SERVS bands but resolved in the fiducial VIDEO band (for this source, $K_{\mathrm{s}}$ band).  The cutout dimensions are $10^{\prime \prime} \times 10^{\prime \prime}$ and the source was modeled using a deVaucouleurs profile.  The left column shows the original image, the center column shows the source model convolved with the PSF of each band, and the right column shows the $\chi^2$ image and $\chi^{2}_{\mathrm{red}}$ value after fitting with {\it The Tractor}.  The colorbar units are in image counts for VIDEO and CFHTLS-D1 and MJy~sr$^{-1}$ for SERVS.}
\label{fig:photometry_example}
\end{figure}
%%%%%%%%%%%%%%%%%%%%%%%%%%%%%%%%%%%%%%%%%%%%%%

%%%%%%%%%%%%%%%%%%%%%%%%%%%%%%%%%%%%%%%%%%%%%%%
\begin{figure}
\centering
\includegraphics[clip=true, trim=0cm 0cm 0cm 0.25cm, height=20cm]{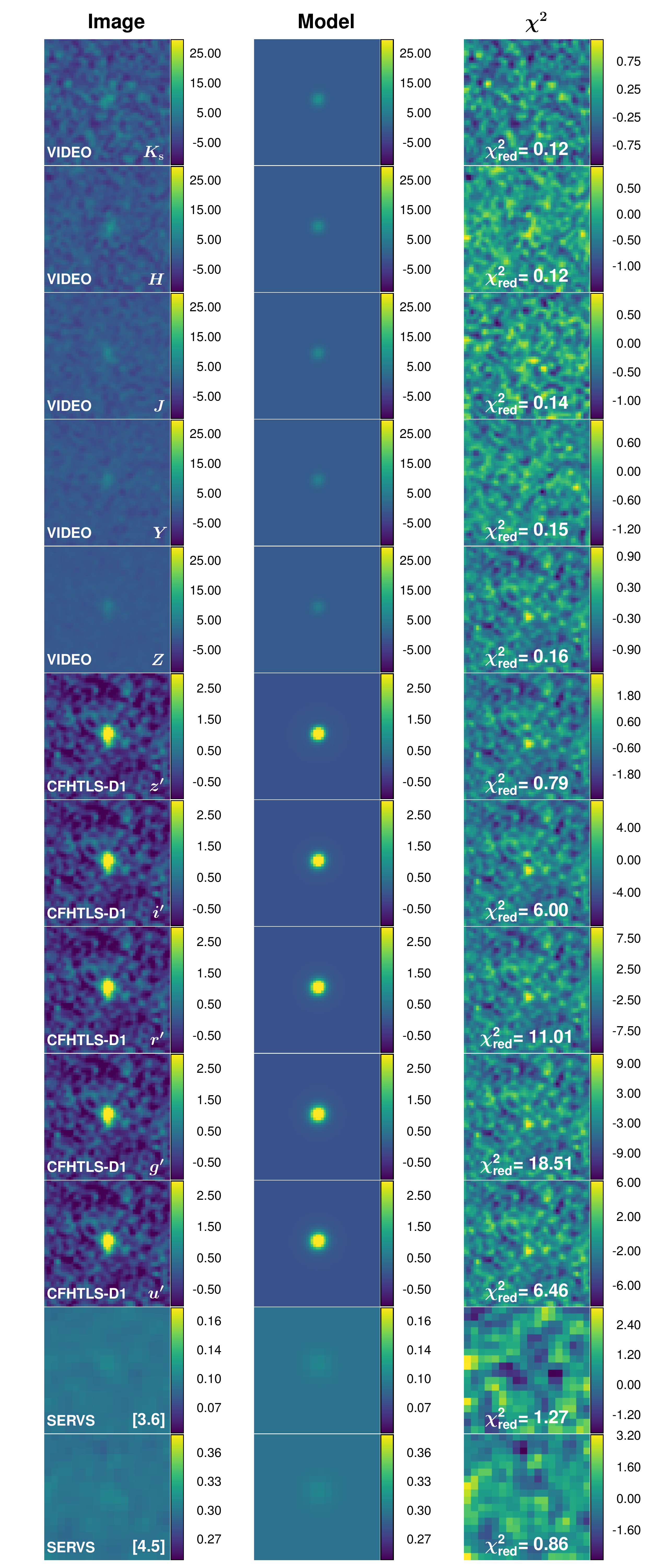}
\caption{Example of our forced photometry procedure for a source with no blending issues that is much fainter than the example shown in Figure~\ref{fig:photometry_example}.  The cutout dimensions are $10^{\prime \prime} \times 10^{\prime \prime}$ and the source was modeled using a point source model.  The left column shows the original image, the center column shows the source model convolved with the PSF of each band, and the right column shows the $\chi^2$ image and $\chi^{2}_{\mathrm{red}}$ value after fitting with {\it The Tractor}.  The colorbar units are in image counts for VIDEO and CFHTLS-D1 and MJy~sr$^{-1}$ for SERVS.}
\label{fig:photometry_example2}
\end{figure}
%%%%%%%%%%%%%%%%%%%%%%%%%%%%%%%%%%%%%%%%%%%%%%

%%%%%%%%%%%%%%%%%%%%%%%%%%%%%%%%%%%%%%%%%%%%%%%%
\subsection{Optimization}
Given a source with the information described above, {\it The Tractor} performs a least-squares fit to the image data to determine the source brightnesses.  While in principle all parameters may be left free to vary (i.e., source positions, shapes, and fluxes), we found that allowing too many parameters to vary caused some fits to yield unphysical results.  To avoid these issues, we held all image and calibration parameters fixed during optimization except for the fluxes, which are left free to vary.  This type of photometric fitting strategy is sometimes referred to as ``forced photometry''\footnote{We note that usage of the term ``forced photometry'' is not consistent throughout the literature.  In some publications, the term refers to the process of smoothing all images to match the lowest resolution band and then performing matched-aperture photometry.  Here, our usage of the term follows from \citet{lang+16} and describes the process of using prior information from a high-resolution band to model the flux at the same position in lower-resolution bands.}.  Examples of the original multi-band images, models, and $\chi^{2}$ maps for a blended IRAC source and a non-blended, faint IRAC source for which {\it The Tractor} has produced improved multi-band photometry compared to the original input catalog is shown in Figures~\ref{fig:photometry_example} and \ref{fig:photometry_example2}.

Our implementation of {\it The Tractor} required the development of a parallelized {\sc Python} driver script.  Parallelization was performed using the Multiprocessing {\sc Python} module.  A full run of our script for the 117,281 sources with imaging available over 12 bands in our VIDEO-selected input catalog took approximately 16 hours on a cluster node with 16 cores and 64 GB of memory.  Diagnostic images (original sub-image cutout, source model image, and $\chi^2$ array) may be optionally produced, though this significantly increases the run time of our code.

%%%%%%%%%%%%%%%%%%%%%%%%%%%%%%%%%%%%%%%%%%%%%%%%%
\subsection{Output Catalogs}
The forced photometry of the VIDEO- and IRAC-selected output catalogs produce multi-band measurements of source fluxes and magnitudes as well as errors.  Information on the source position, fiducial band, best-fitting surface brightness model, and $\chi^{2}_{\mathrm{red}}$ value after fitting with {\it The Tractor} is also included in the output catalogs.  We provide these catalogs in the online supplementary information, and present additional details on their contents in the appendix.  

%%%%%%%%%%%%%%%%%%%%%%%%%%%%%%%%%%%%%%%%%%%%%%%%%
\subsubsection{Saturated Sources}
\begin{figure*}[h!]
\centering
\includegraphics[clip=true, trim=0cm 0.2cm 0cm 0.2cm, width=5.0cm]{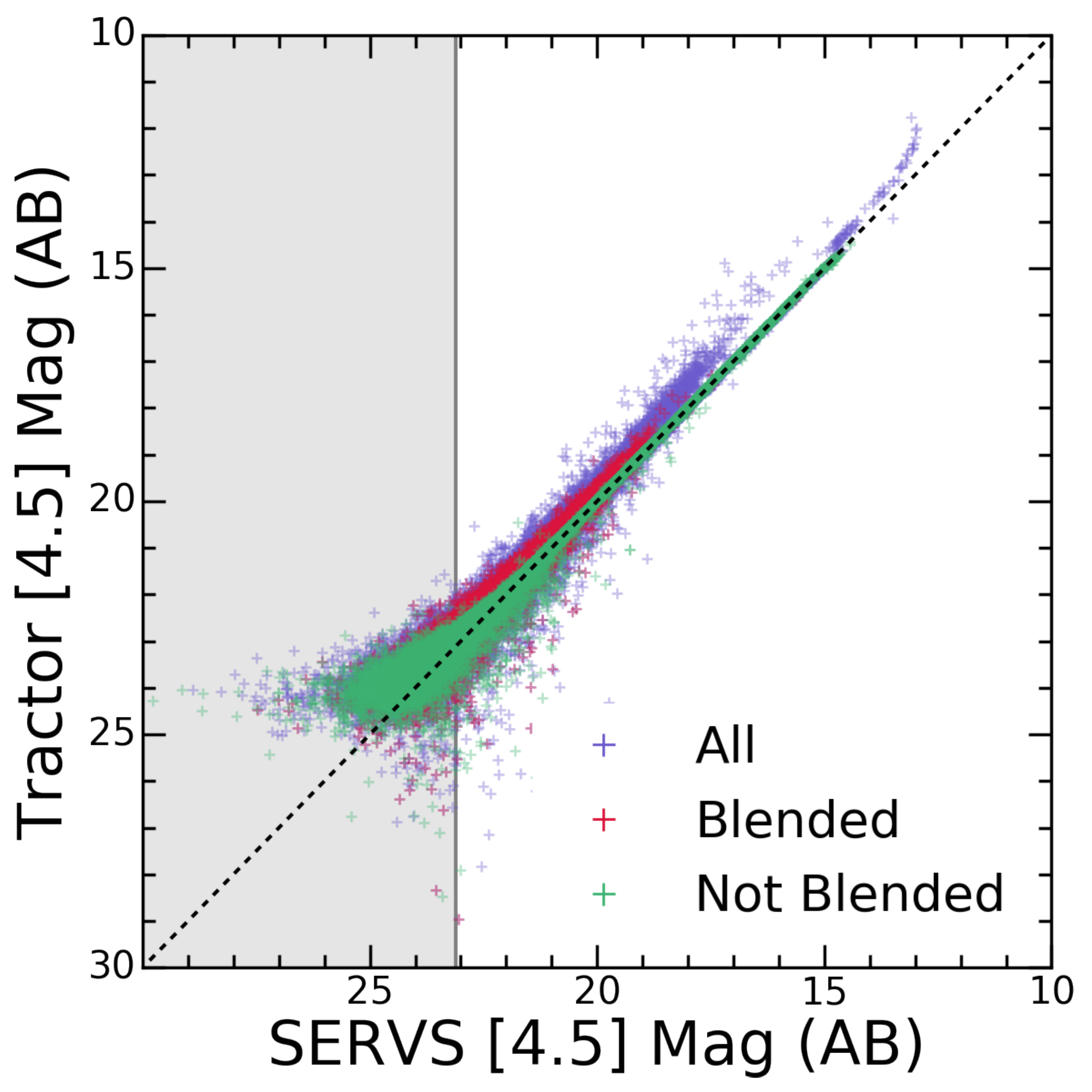}
\includegraphics[clip=true, trim=0cm 0.2cm 0cm 0.2cm, width=5.0cm]{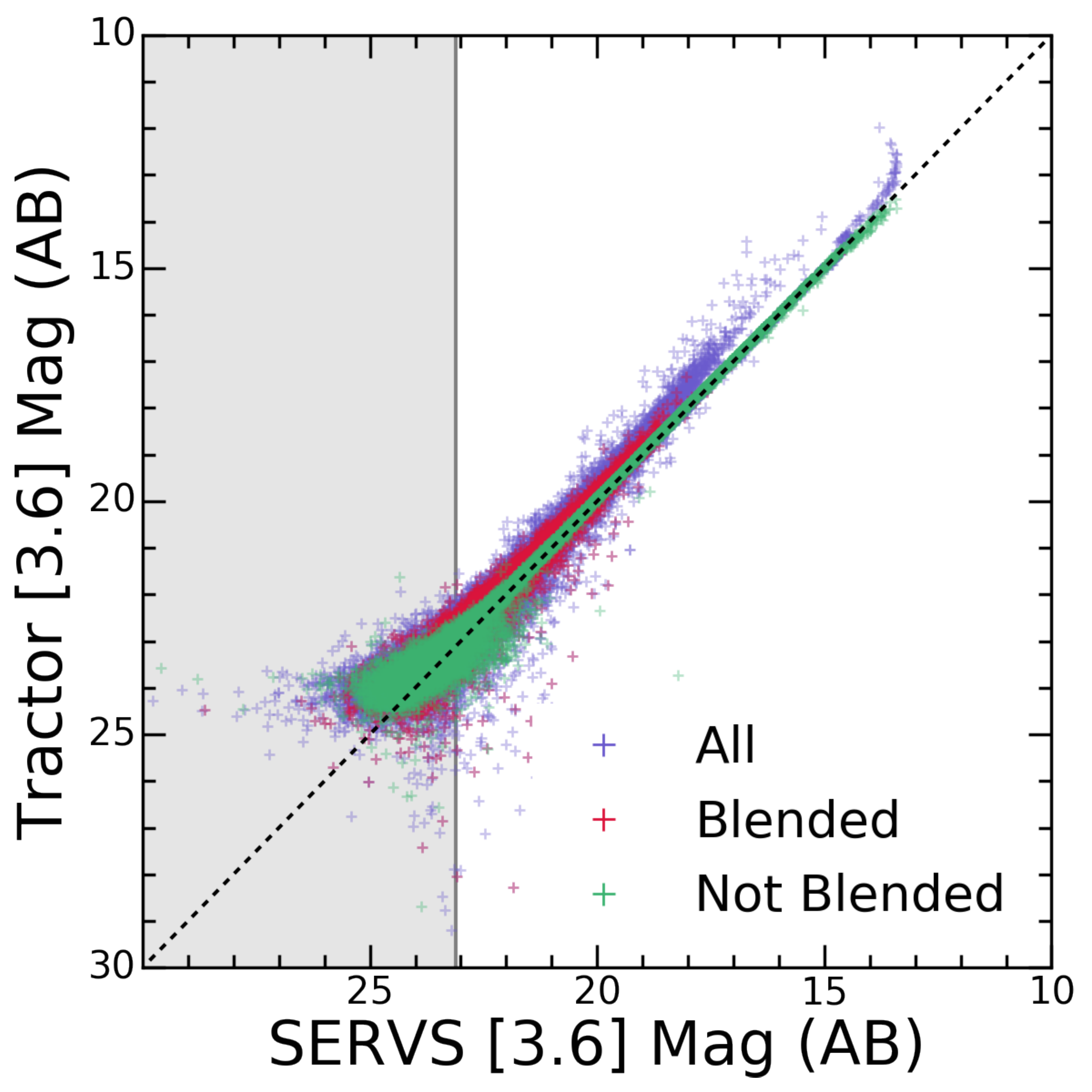}
\includegraphics[clip=true, trim=0cm 0.2cm 0cm 0.2cm, width=5.0cm]{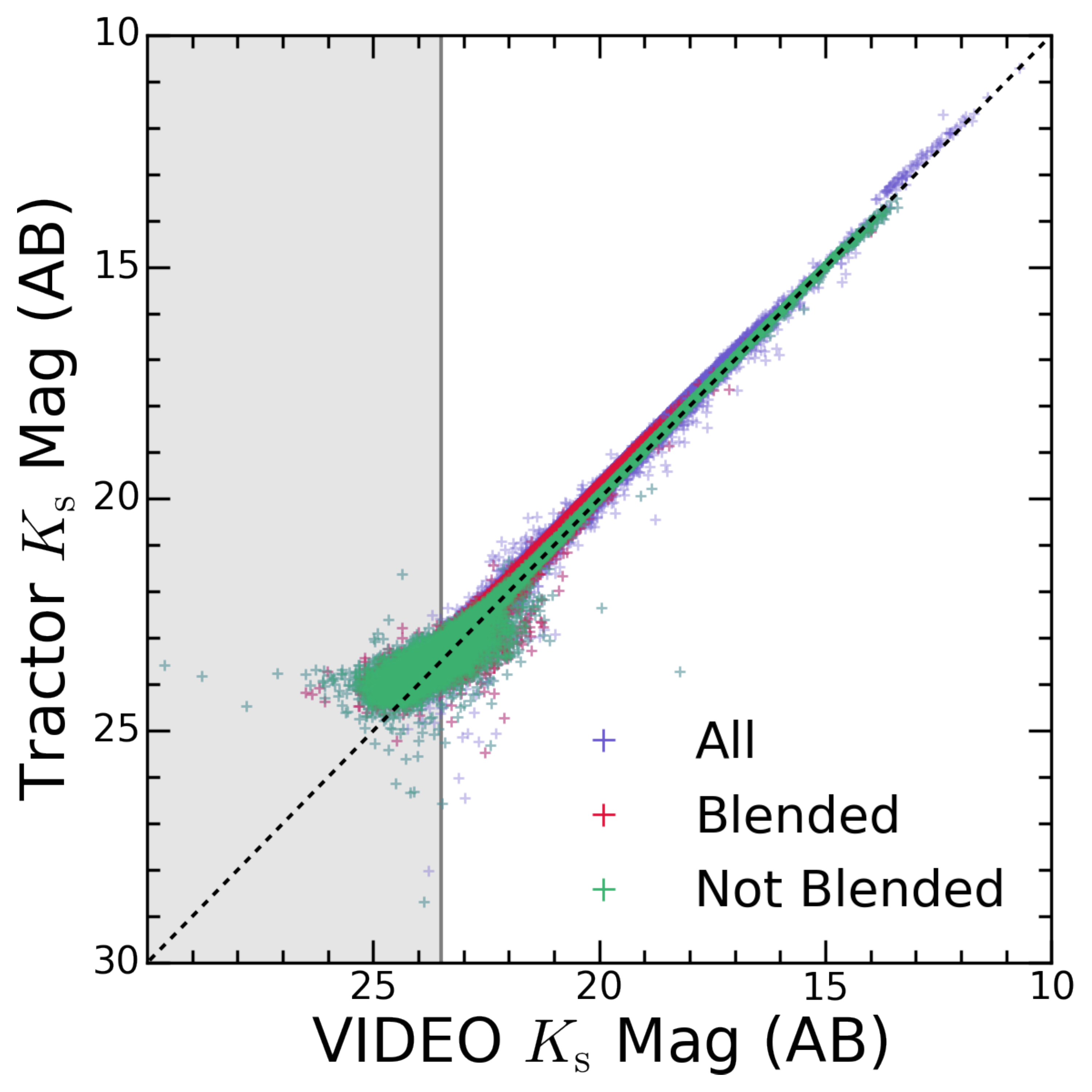}

\includegraphics[clip=true, trim=0cm 0.2cm 0cm 0.2cm, width=5.0cm]{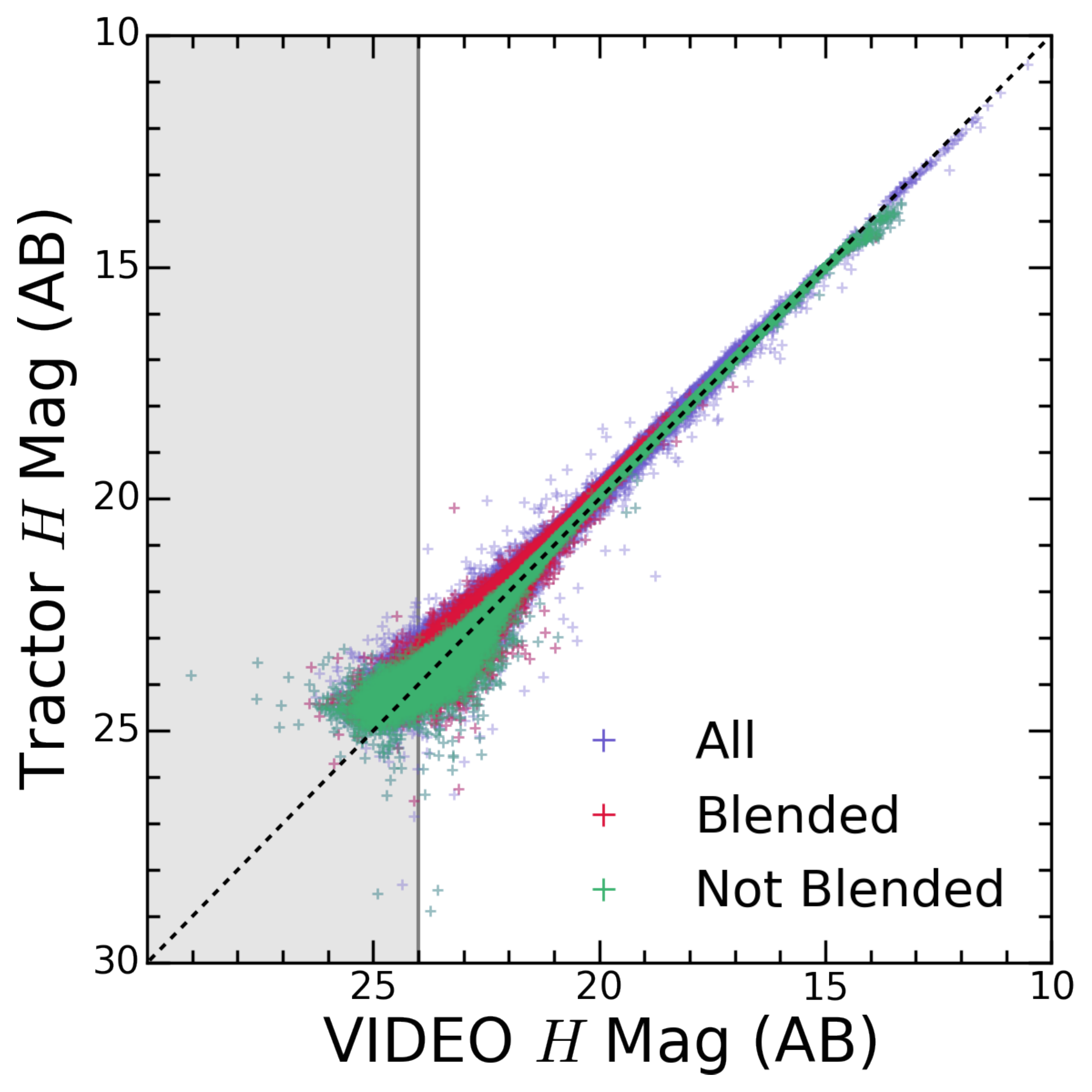}
\includegraphics[clip=true, trim=0cm 0.2cm 0cm 0.2cm, width=5.0cm]{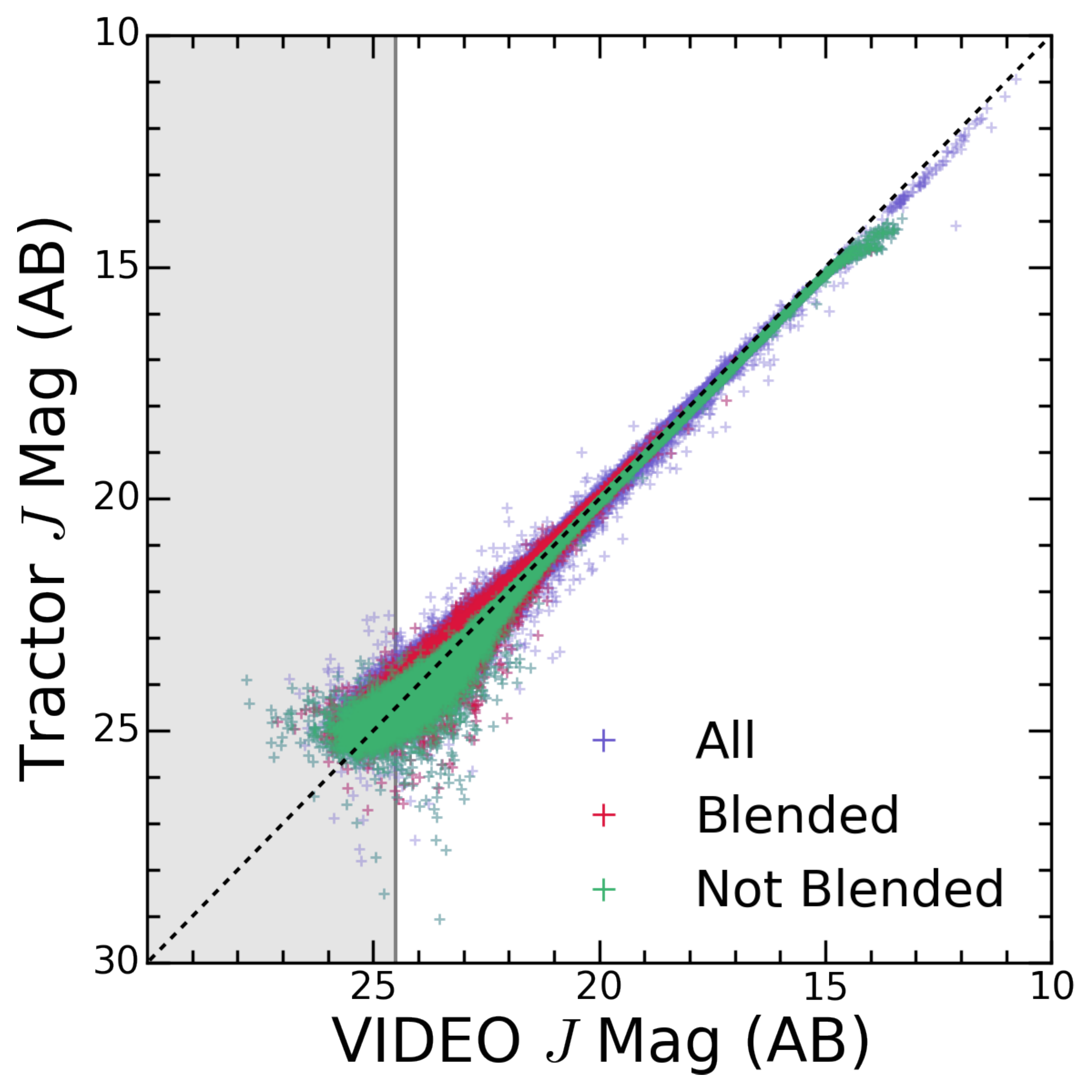}
\includegraphics[clip=true, trim=0cm 0cm 0cm 0.2cm, width=5.0cm]{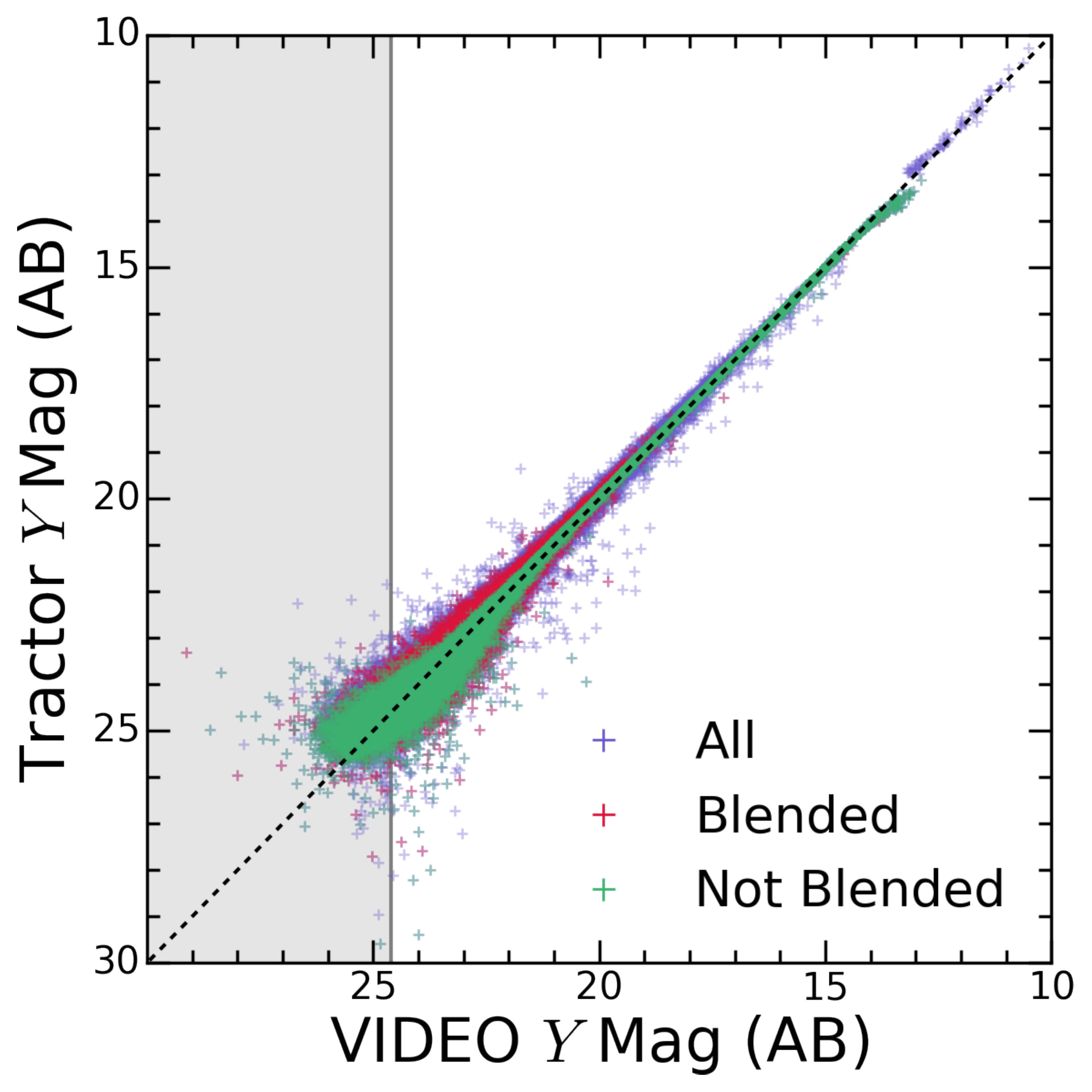}

\includegraphics[clip=true, trim=0cm 0.2cm 0cm 0.2cm, width=5.0cm]{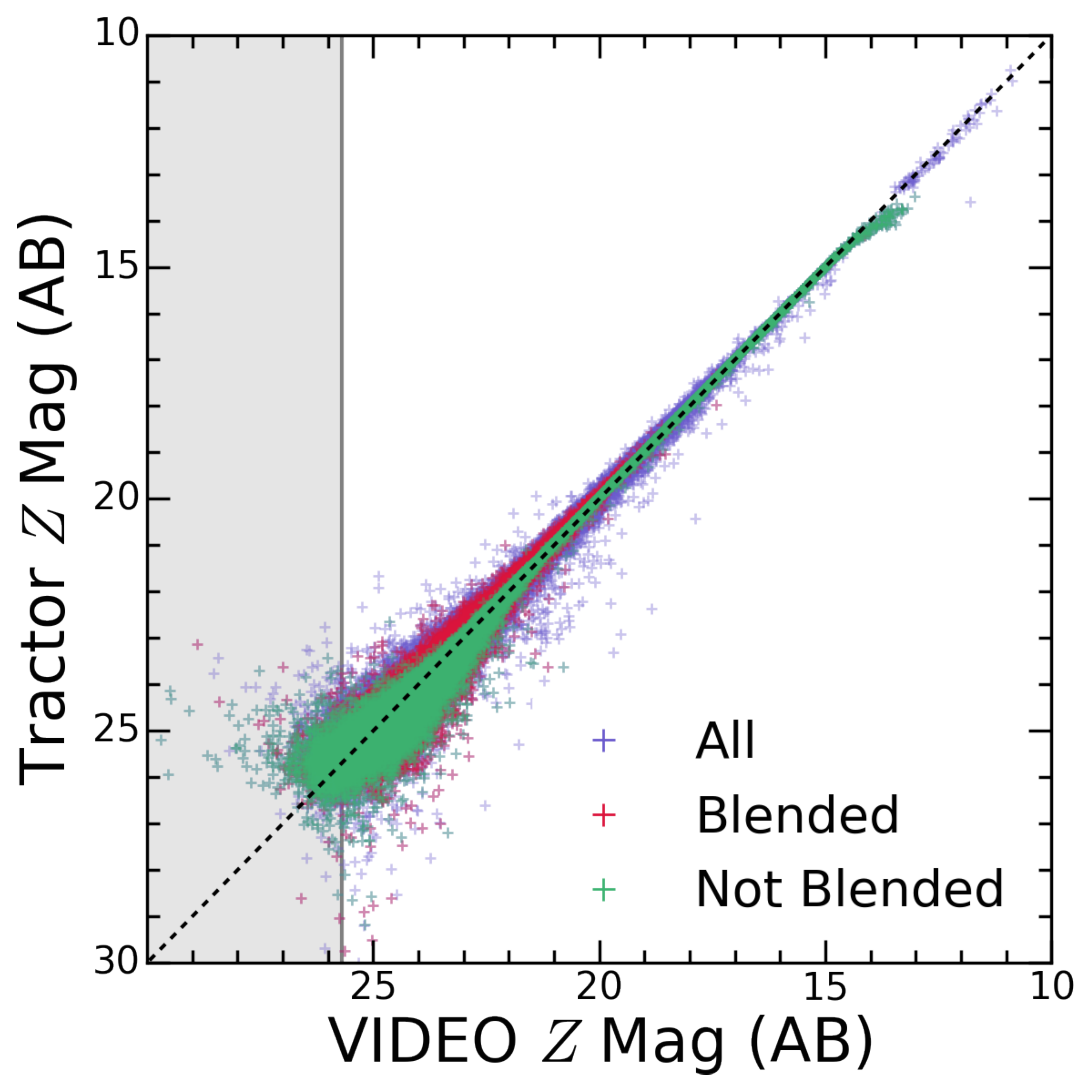}
\includegraphics[clip=true, trim=0cm 0.2cm 0cm 0.2cm, width=5.0cm]{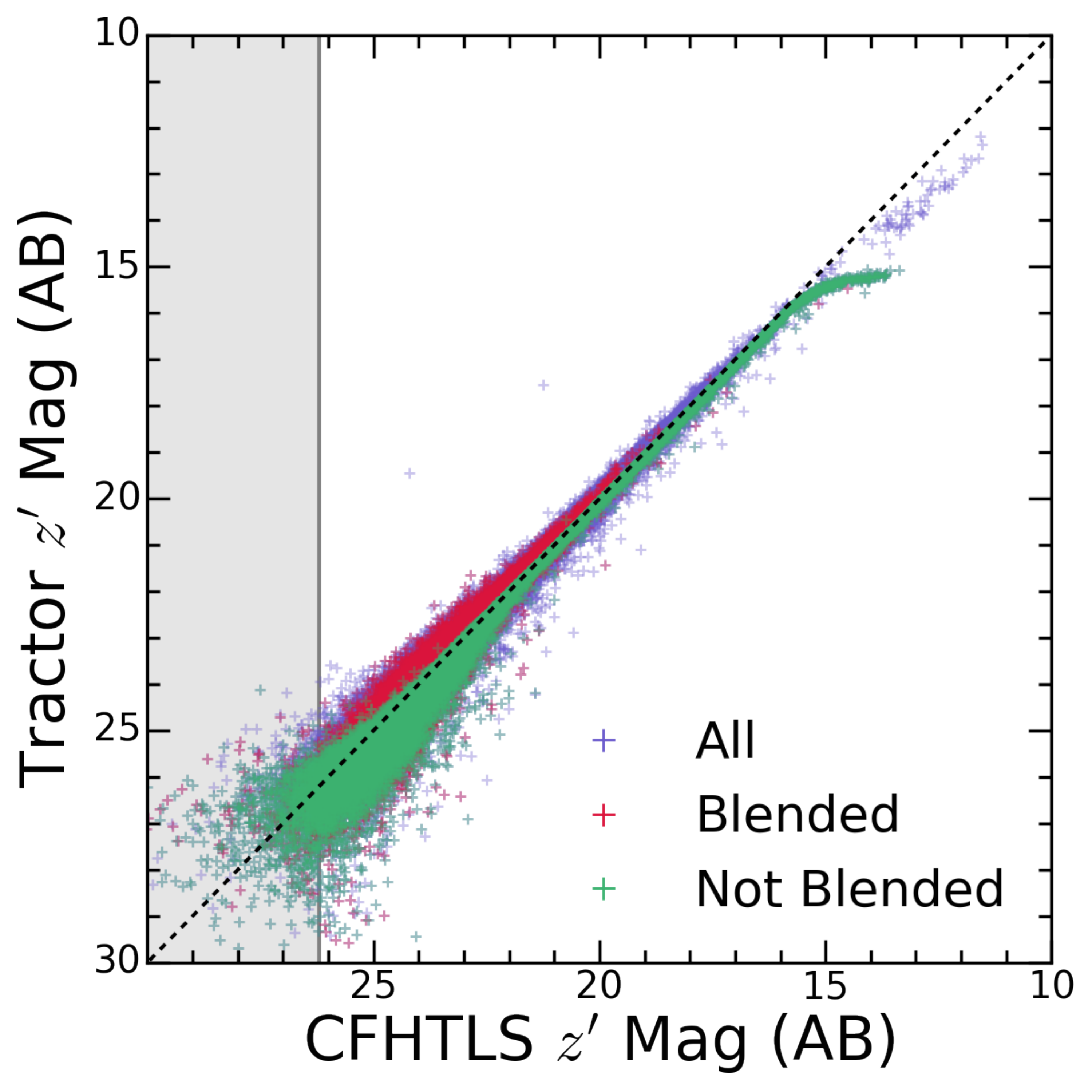}
\includegraphics[clip=true, trim=0cm 0.2cm 0cm 0.2cm, width=5.0cm]{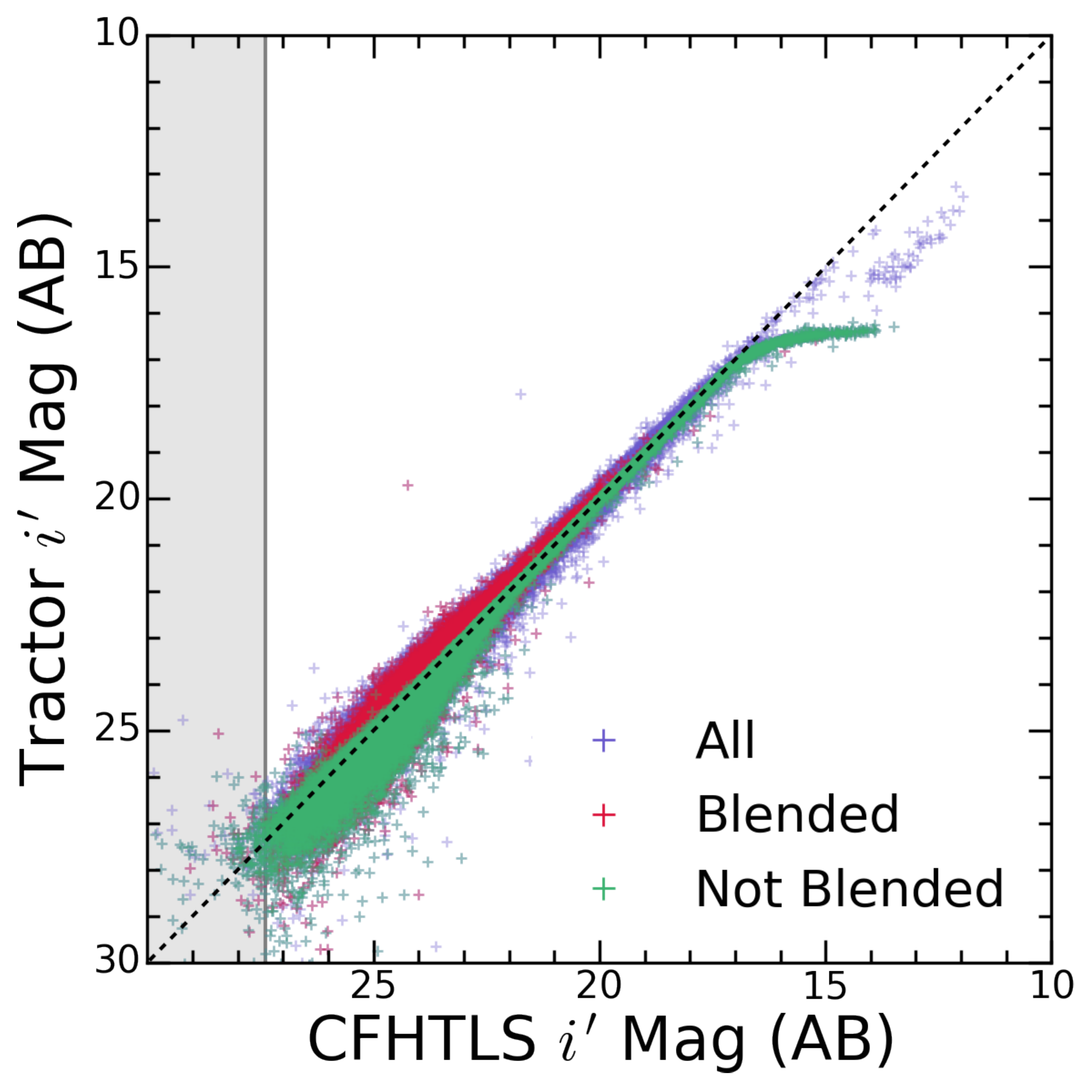}

\includegraphics[clip=true, trim=0cm 0.2cm 0cm 0.2cm, width=5.0cm]{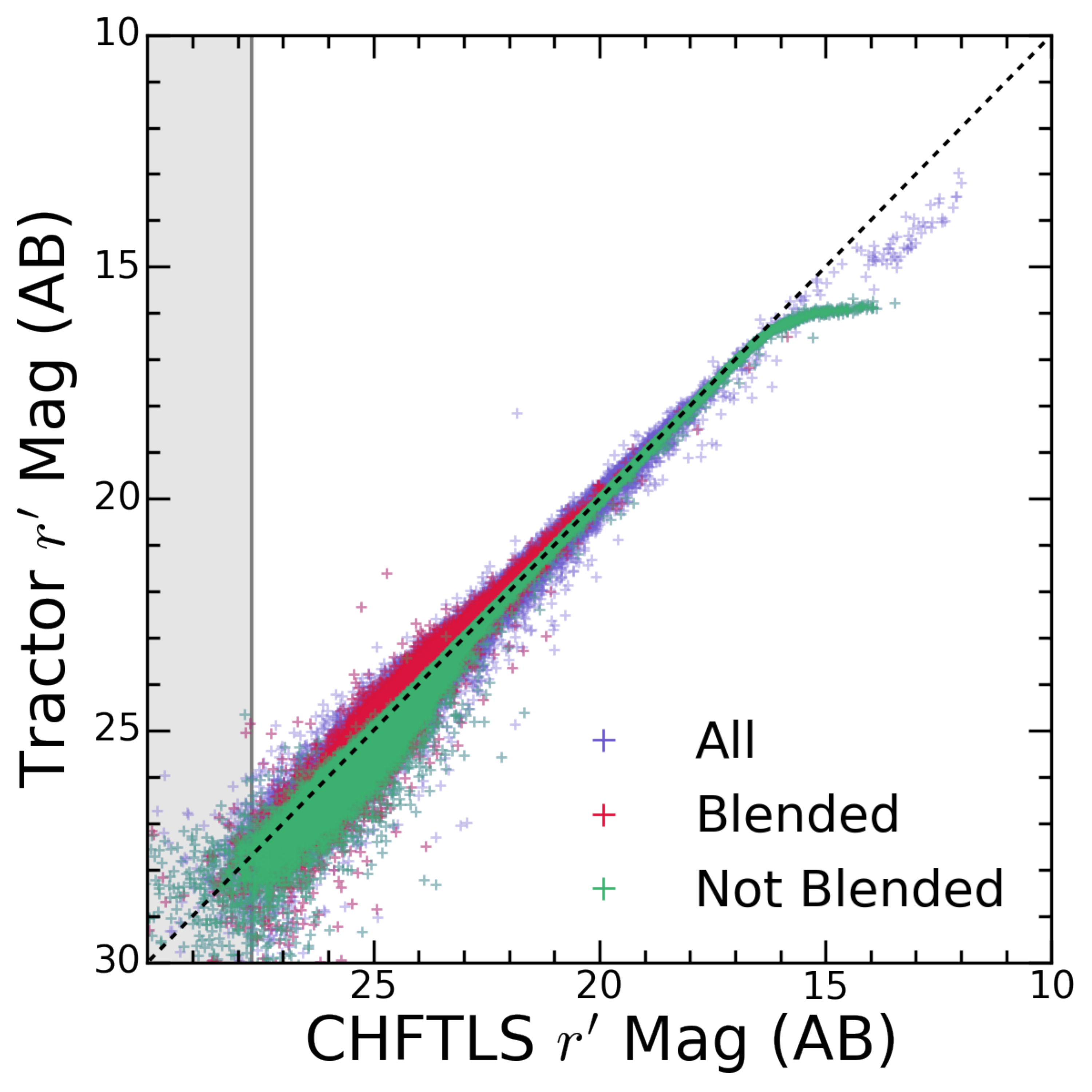}
\includegraphics[clip=true, trim=0cm 0.2cm 0cm 0.2cm, width=5.0cm]{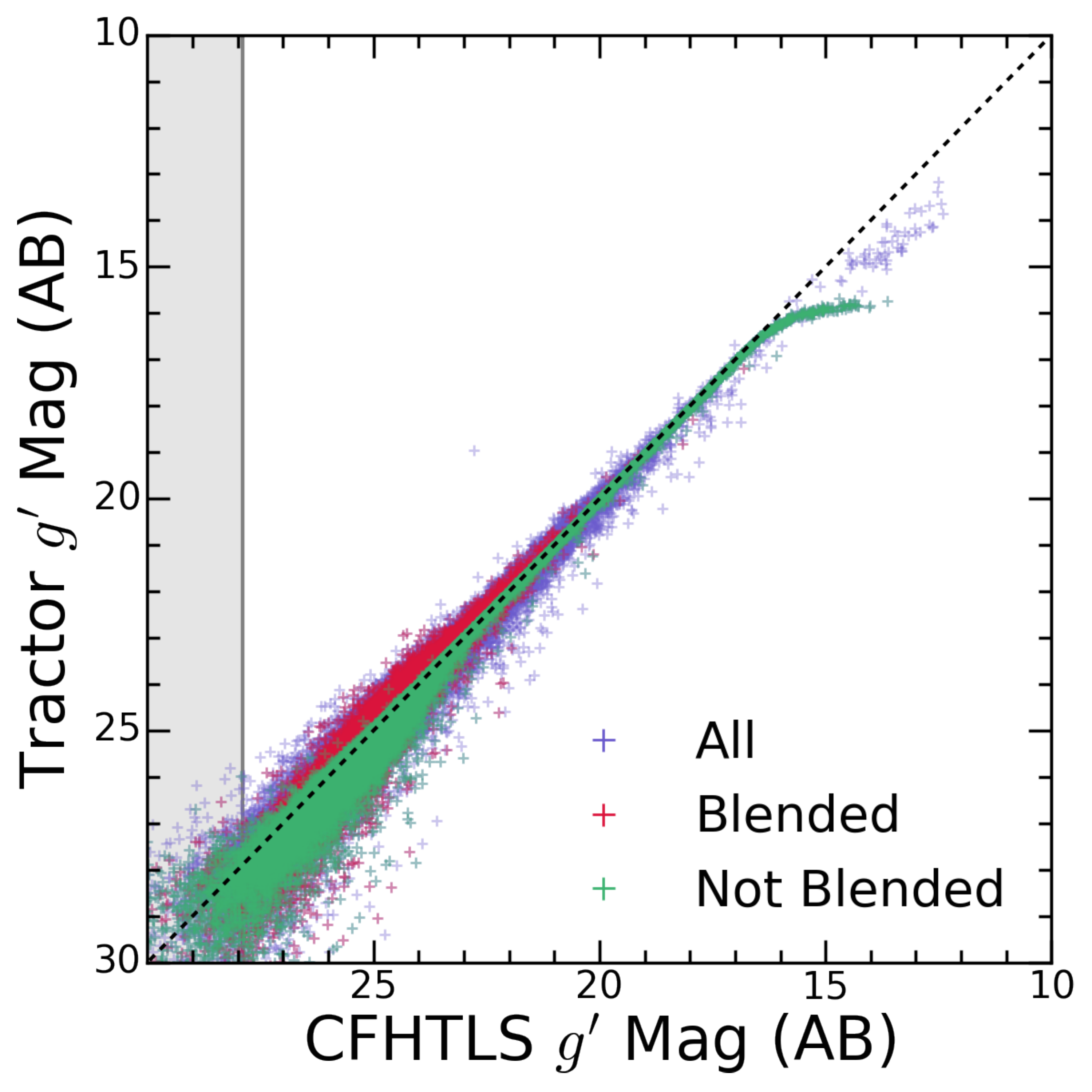}
\includegraphics[clip=true, trim=0cm 0.2cm 0cm 0.2cm, width=5.0cm]{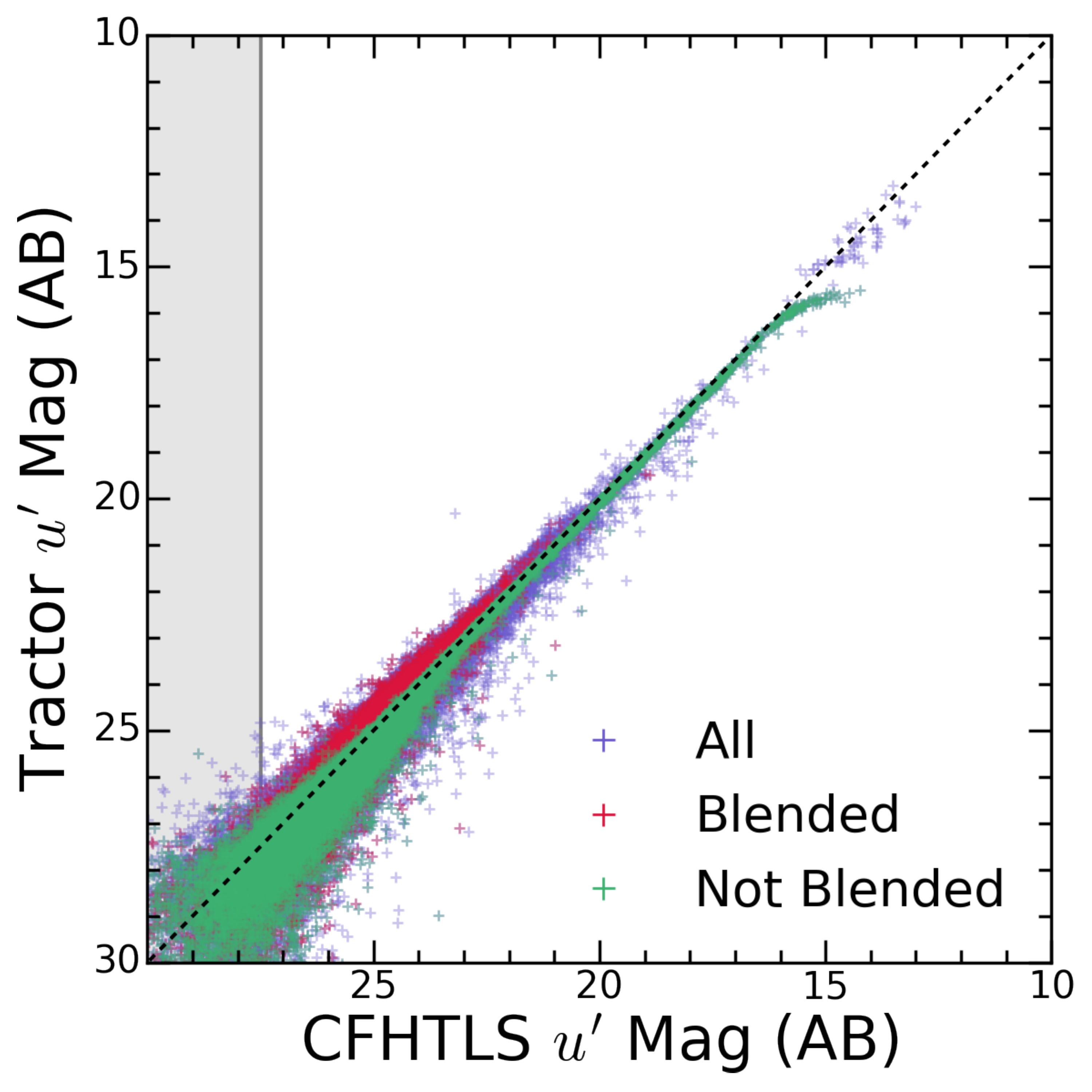}

\caption{Comparison between {\it The Tractor} and original photometry.  For SERVS, we converted the aperture-corrected fluxes measured within an aperture of radius $1\farcs9$ from the original catalog to AB magnitudes.  For VIDEO, we show the Petrosian magnitudes.  For CFHTLS-D1, we show the MAG\_AUTO magnitudes, which are measured within an elliptical aperture similar to that defined in \citet{kron+80}.  The dashed line shows the one-to-one correspondence between {\it The Tractor} and original catalog magnitudes.  Blended sources in SERVS identified based on the presence of a nearby source in the VIDEO catalog within $3\farcs8$ are shown in red.  Sources lacking neighbors in VIDEO within $3\farcs8$ that were modeled as point sources and, therefore, known to be free of blending issues in SERVS are shown in green.  The purple symbols trace all sources.  This includes clearly blended/non-blended sources and sources that were modeled with a resolved surface brightness profile.  The gray-shaded region highlights the parameter space below the average 5$\sigma$ detection threshold of each survey. 
%(KSPETROMAG, HPETROMAG, JPETROMAG, YPETROMAG, and ZPETROMAG)
}
\label{fig:cat_trac_mag_compare_all}
\end{figure*}

We note that some of the brightest sources in the VIDEO-selected output catalog may be saturated.  Thus, we suggest that users wishing to avoid the inclusion of such sources in subsequent analyses utilizing our {\it Tractor} VIDEO-selected output catalog consider the binary saturation flag we have provided in the catalog.  The saturation flag column identifies sources with a high probability of being saturated based on the comparison between {\it The Tractor} and original photometry shown in Figure~\ref{fig:cat_trac_mag_compare_all}.  For the VIDEO data, sources with magnitudes brighter than 14.0 for $Ks$ and $H$ band, as well as sources brighter than 14.5, 13.6, and 13.8 for the $J$, $Y$, and $Z$ bands, respectively, are flagged as saturated.  For the CFHTLS-D1 data, sources brighter than 16.3, 15.9, 15.9, 15.1, and 15.7 for the $i^{\prime}$, $r^{\prime}$, $g^{\prime}$, $z^{\prime}$, and $u^{\prime}$ bands, respectively, are flagged.  Finally, sources brighter than magnitude 14.0 for the IRAC 3.6~$\mu$m band and 13.5 for the IRAC 4.5~$\mu$m band are flagged.  For further details, we refer readers to Table~\ref{tab:cat_columns} in the appendix.  

%%%%%%%%%%%%%%%%%%%%%%%%%%%%%%%%%%%%%%%%%%%%%%%%
%%%%%%%%%%%%%%%%%%%%%%%%%%%%%%%%%%%%%%%%%%%%%%%%
%%%%%%%%%%%%%%%%%%%%%%%%%%%%%%%%%%%%%%%%%%%%%%%

%%%%%%%%%%%%%%%%%%%%%%%%%%%%%%%%%%%%%%%%%%%%%%%%%
\begin{deluxetable*}{cCCCCCC}[t!]
\tablecaption{Median Photometric Offsets \label{tab:cat_offsets}}
\tablecolumns{7}
%\tablenum{3}
\tablewidth{0pt}
\tablehead{
\colhead{Band} & \colhead{$N_{\mathrm{All}}$} & \colhead{$\Delta M_{\mathrm{All}}$} & \colhead{$N_{\mathrm{Blended}}$}  & \colhead{$\Delta M_{\mathrm{Blended}}$} & \colhead{$N_{\mathrm{Not \, \, Blended}}$}  & \colhead{$\Delta M_{\mathrm{Not \, \, Blended}}$}\\
\colhead{(1)} & \colhead{(2)} & \colhead{(3)} & \colhead{(4)} & \colhead{(5)} & \colhead{(6)} & \colhead{(7)}
}
\startdata
Near Infrared & & & & & & \\ 
\hline
4.5$\mu$m           &           99009  &      -0.218  &           15730  &      -0.155  &                25683  &           -0.043  \\ 
3.6$\mu$m           &          103911  &      -0.235  &           15808  &      -0.175  &                28657  &           -0.097  \\ 
$K_{\mathrm{s}}$ &           98811  &      -0.230  &           15928  &      -0.215  &                23755  &           -0.111  \\ 
$H$                &          104752  &      -0.176  &           17129  &      -0.154  &                25598  &           -0.035  \\ 
$J$                 &          106733  &      -0.006  &           17473  &       0.020  &                26036  &            0.118  \\ 
$Y$                 &           99516  &      -0.113  &           15991  &      -0.081  &                23390  &           -0.014  \\ 
$Z$                 &          107651  &      -0.063  &           17714  &      -0.013  &                28601  &            0.040  \\ 
\hline
Optical & & & & & &  \\ 
\hline
$z^{\prime}$    &     104891  &      -0.138  &           17286  &      -0.008  &                29651  &            0.233  \\ 
$i^{\prime}$     &  105392  &      -0.008  &           17332  &      -0.009  &                29841  &            0.224  \\ 
$r^{\prime}$     &  105044  &       0.006  &           17256  &       0.004  &                29714  &            0.198  \\ 
$g^{\prime}$    &  104124  &       0.020  &           17093  &       0.019  &                29275  &            0.172  \\ 
$u^{\prime}$    &  98416  &       0.051  &           16081  &       0.045  &                26688  &            0.202  \\ 
\enddata
\tablecomments{Column 1: Observing band or filter name.  Column 2:  The number of sources with photometric measurements available in both the original VIDEO-selected input catalog and our new multi-band forced photometric catalog.  Column 3: The median difference in magnitude between our new forced photometry with {\it The Tractor} and the original catalog photometry.  Column 4: The number of sources known to be blended in the SERVS catalog based on the presence of at least one neighboring source within $3\farcs8$ in the VIDEO catalog.  Column 5: Same as Column 3, except the median magnitude difference is calculated for blended sources only.  Column 6: The number of isolated sources lacking neighbors within $3\farcs8$ that were modeled as point sources in our forced photometry and are thus not expected to have any blending issues in the SERVS images.  Column 7: Same as Column 3, except here the median magnitude difference is calculated for sources that are not expected to be blended.}
\end{deluxetable*}
%%%%%%%%%%%%%%%%%%%%%%%%%%%%%%%%%%%%%%%%%%%%%%%%%%

%%%%%%%%%%%%%%%%%%%%%%%%%%%%%%%%%%%%%%%%%%%%%%%%
\subsection{Caveats}
Although our new multi-band photometric catalogs produced using {\it The Tractor} offer important advantages over existing catalogs for blended and/or intrinsically faint sources,  there are a number of important caveats.  We emphasize that improved photometry of blended IRAC sources can only be achieved if the blended objects are well-resolved in the fiducial VIDEO band used to generate the source model.  For highly complex, extended sources not well-described by a deVaucouleurs or exponential model, the accuracy of our photometry with {\it The Tractor} will be reduced.  The inclusion of additional surface brightness profile models and/or performing fitting with {\it The Tractor} over multiple iterations may help address this issue in the future.  We note that our strategy assumes that the source surface brightness profile is the same at all 12 NIR and optical bands included in our analysis.  In other words, we effectively assume morphological $k$ corrections are small.  
%Thus, our strategy for implementing forced photometry described in this study can only be applied to multi-band datasets spanning a narrow enough range in wavelength such that the assumption of small morphological $k$ corrections is reasonable.

Our photometry also does not take into account spatial variations in the PSF or sky background level, which could lead to aperture errors that are difficult to correct and poorer flux measurement accuracy for fainter sources, respectively.  While in principle it would be possible to provide {\it The Tractor} with position-dependent PSF information for all bands based on models generated using the {\sc PSFEx} software \citep{bertin+11}, this would increase the computational cost substantially.

Although position-dependent astrometric variations can in principle compromise the photometric accuracy of {\it The Tractor}, the datasets used here do not suffer significantly from such effects.  Both VIDEO and CFHTLS-D1 have relative astrometric uncertainties $< 0\farcs1$ \citep{jarvis+13, gwyn+12}.  For the SERVS data, the IRAC Instrument Handbook\footnote{http://irsa.ipac.caltech.edu/data/SPITZER/docs/irac} reports that the astrometry is typically accurate to $\sim 0\farcs2$, or about the size of a single pixel in the VIDEO survey.  Given these relatively small uncertainties, we don't expect astrometric errors to be a dominant limiting factor in the accuracy of our forced photometry.

%%%%%%%%%%%%%%%%%%%%%%%%%%%%%%%%%%%%%%%%%%%%%%%%
%%%%%%%%%%%%%%%%%% DISCUSSION %%%%%%%%%%%%%%%%%%%%%%%
%%%%%%%%%%%%%%%%%%%%%%%%%%%%%%%%%%%%%%%%%%%%%%%%
\section{Results}
\label{sec:results}

%%%%%%%%%%%%%%%%%%%%%%%%%%%%%%%%%%%%%%%%%%%%%%%%
\subsection{VIDEO-Selected Catalog}
\label{sec:vid_sel_cat}
We find that about 65\% of the sources in the VIDEO-selected forced photometry catalog are extended based on the criteria described in Section~\ref{sec:surf_bright_prof}, and require spatially-resolved surface brightness profile models.  The high fraction of extended sources suggests that the number of blended SERVS sources is indeed significantly higher than our lower limit of 17\% (see Section~\ref{sec:input_cats}).  Of the resolved sources, the majority are best fit by a deVaucouleurs profile ($\sim$61\%) rather than an exponential profile ($\sim$39\%).  The vast majority ($\sim$84\%) of the sources were modeled using the VIDEO $K_{\mathrm{s}}$-band data as the fiducial band.

A comparison of the source magnitudes from {\it The Tractor} forced photometry and the original photometry for the VIDEO-selected catalog is shown in Figure~\ref{fig:cat_trac_mag_compare_all}.  Our forced photometry is typically in good agreement with the original catalog magnitudes, though some scatter is apparent.  As expected, the scatter is largest for faint sources.  The photometry of these sources is likely more sensitive to noise fluctuations across the image that have not been accounted for by our constant noise assumption.  Other factors that may contribute to the scatter include the presence of blended sources, spatial PSF variations, inaccurately matched sources in the input catalog, and issues with the photometry from the original catalogs.  
Restricting the comparison to ``isolated'' sources that lack a neighbor within $3\farcs8$ in the VIDEO-selected input catalog and were also fit with point source models in the {\it Tractor} forced photometry further reduces the scatter in Figure~\ref{fig:cat_trac_mag_compare_all}.  

Table~\ref{tab:cat_offsets} summarizes the median difference between our new forced photometry and the original photometry at each band for all sources, blended sources, and isolated point sources (non-blended sources).  The magnitudes of the offsets between {\it The Tractor} and original catalogs are dominated by sources at or below the original detection thresholds of the respective surveys, a regime in which accurate source brightness measurement is difficult.  The uncertainty in the brightnesses of these faint sources is indicated by their large errors in our output multi-band catalog as well as in the original VIDEO, CFHTLS-D1, and SERVS catalogs.

%%%%%%%%%%%%%%%%%%%%%%%%%%%%%%%%%%%%%%%%%%%%%%%%
\subsection{IRAC-Selected Catalog}
\begin{figure}
\centering
\includegraphics[clip=true, trim=0cm 0cm 0cm 0cm, width=7.8cm]{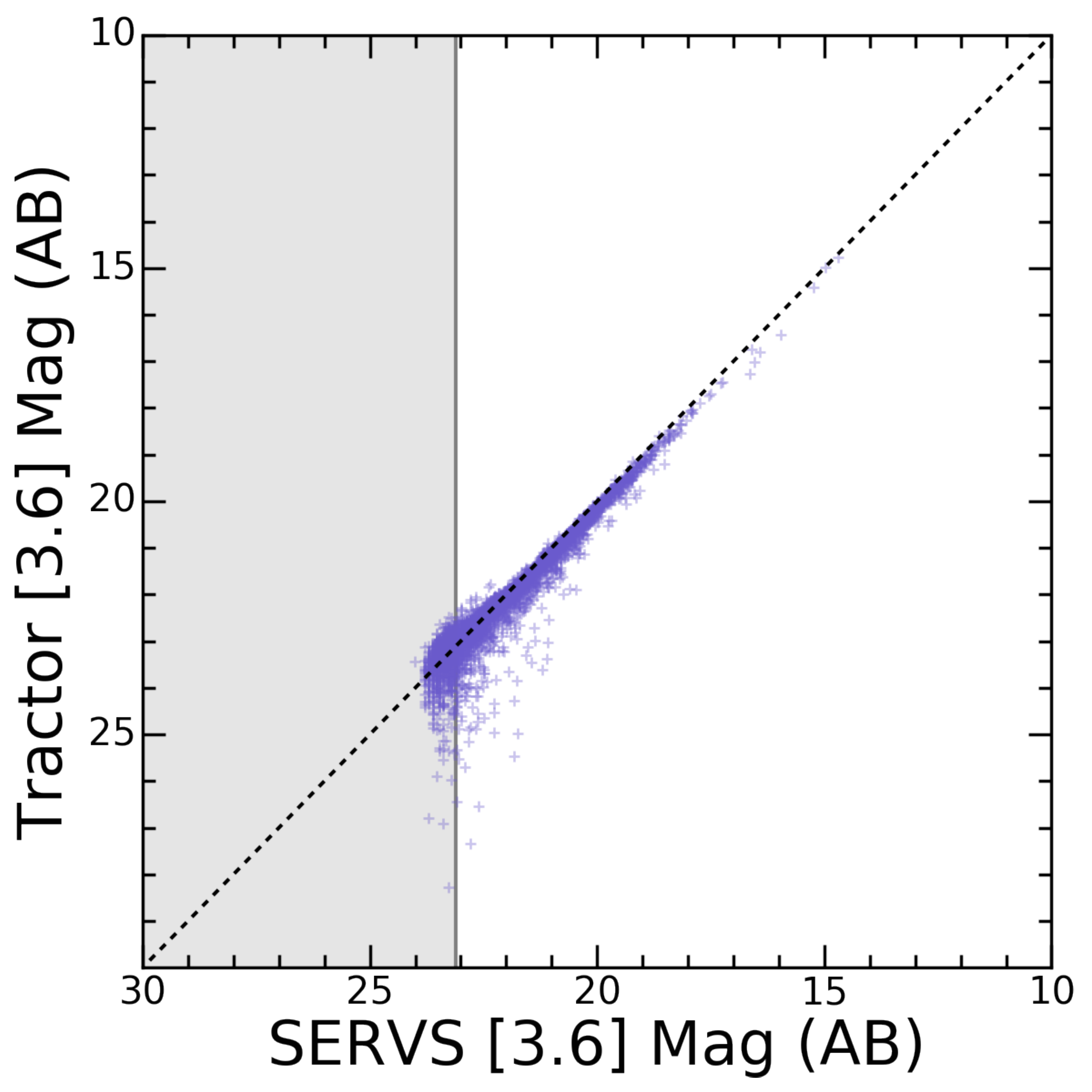}

\includegraphics[clip=true, trim=0cm 0cm 0cm 0cm, width=7.8cm]{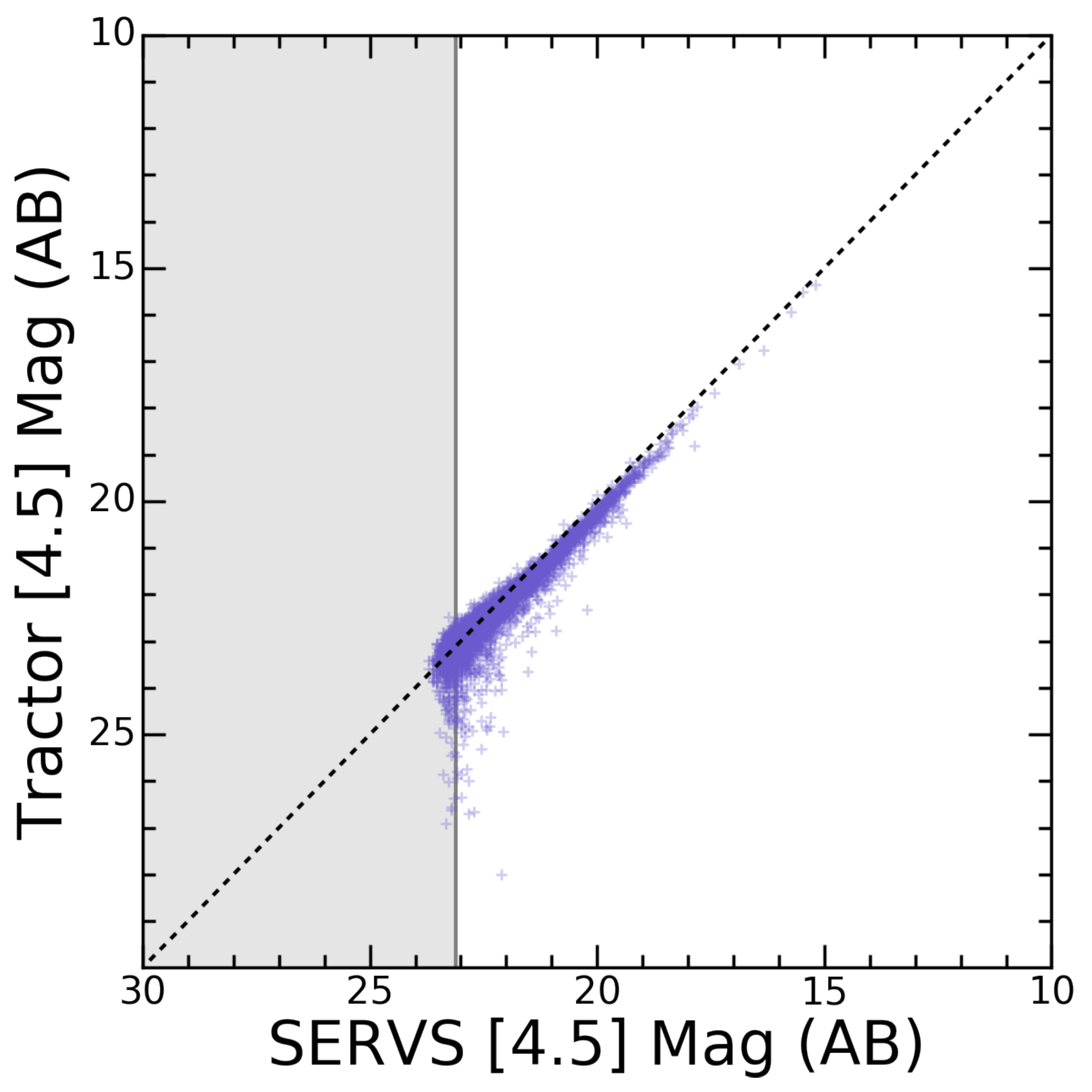}
\centering
\caption{Comparison of our new forced photometry and the IRAC-selected input catalog for the [3.6] (top) and [4.5] (bottom) SERVS bands.  We use the original SERVS single-band aperture-corrected catalog photometry measured within an aperture of radius $1\farcs9$ and converted to AB magnitudes.  The gray-shaded region highlights the parameter space below the average 5$\sigma$ detection threshold of the SERVS data.  The $x$- and $y$-axis data ranges match those from the VIDEO-selected catalog comparison plots shown in Figure~\ref{fig:cat_trac_mag_compare_all}.\\}
\label{fig:cat_trac_mag_diff_compare_irac_only}
\end{figure}

In Figure~\ref{fig:cat_trac_mag_diff_compare_irac_only}, we compare the SERVS photometry from the IRAC-selected input catalog with the results of our new forced photometry.  The photometry in each of these catalogs is generally in good agreement, except for a small population of very faint sources near the SERVS detection limit where the scatter increases notably.  These sources may be extended and characterized by inherently low surface brightness emission, making our assumption of a point source surface brightness profile inadequate.  It is also possible that the original photometry overestimated the source brightness for many of these objects by erroneously including noise and/or emission from nearby confusing sources.

%%%%%%%%%%%%%%%%%%%%%%%%%%%%%%%%%%%%%%%%%%%%%%%%
%%%%%%%%%%%%%%%%%%%%%%%%%%%%%%%%%%%%%%%%%%%%%%%

%%%%%%%%%%%%%%%%%%%%%%%%%%%%%%%%%%%%%%%%%%%%%%%%
\begin{figure*}
\centering
\includegraphics[clip=true, trim=0.5cm 0cm 0cm 0cm, width=8.75cm]{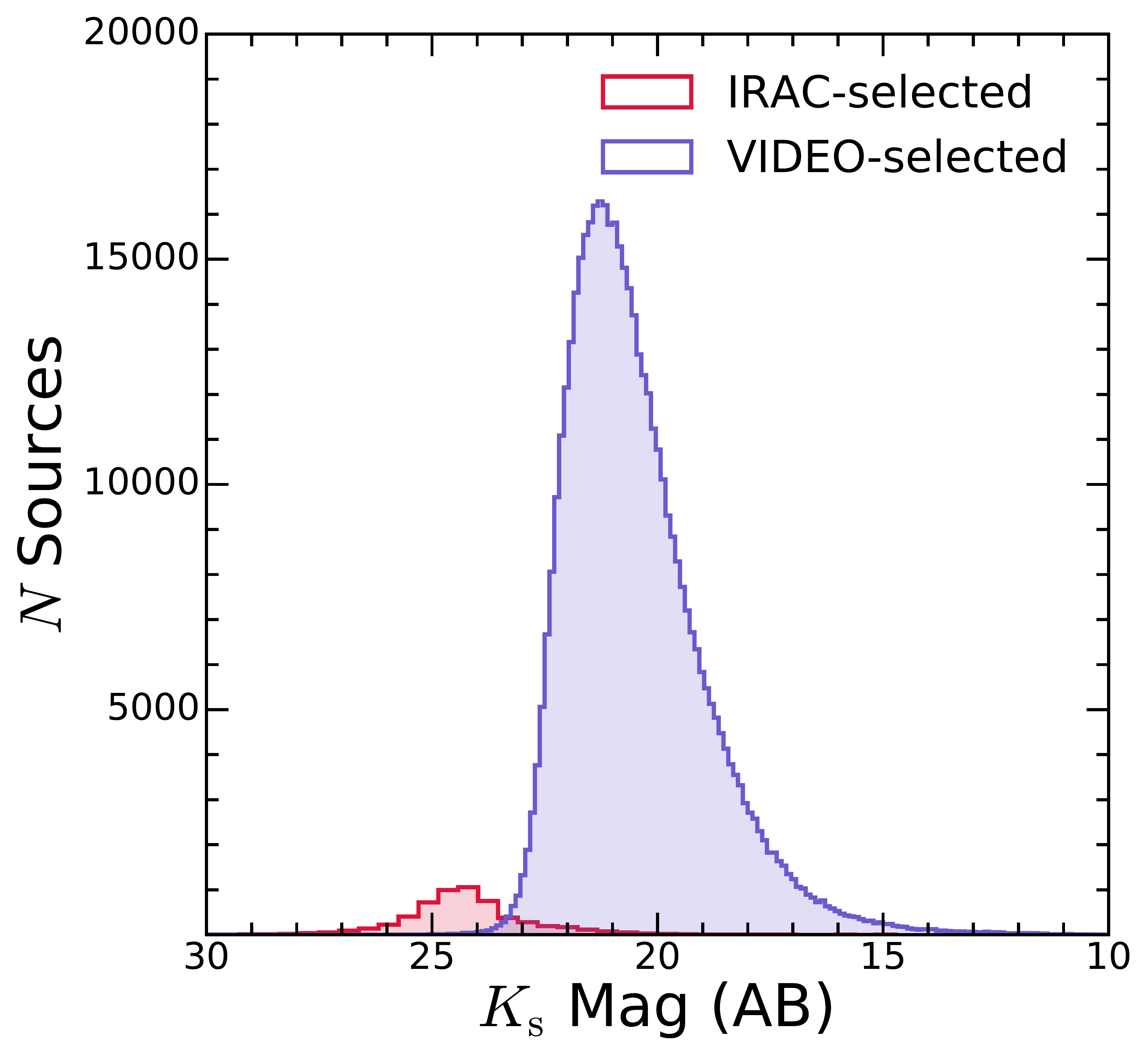}
\includegraphics[clip=true, trim=0.5cm 0cm 0.5cm 0cm, width=8.75cm]{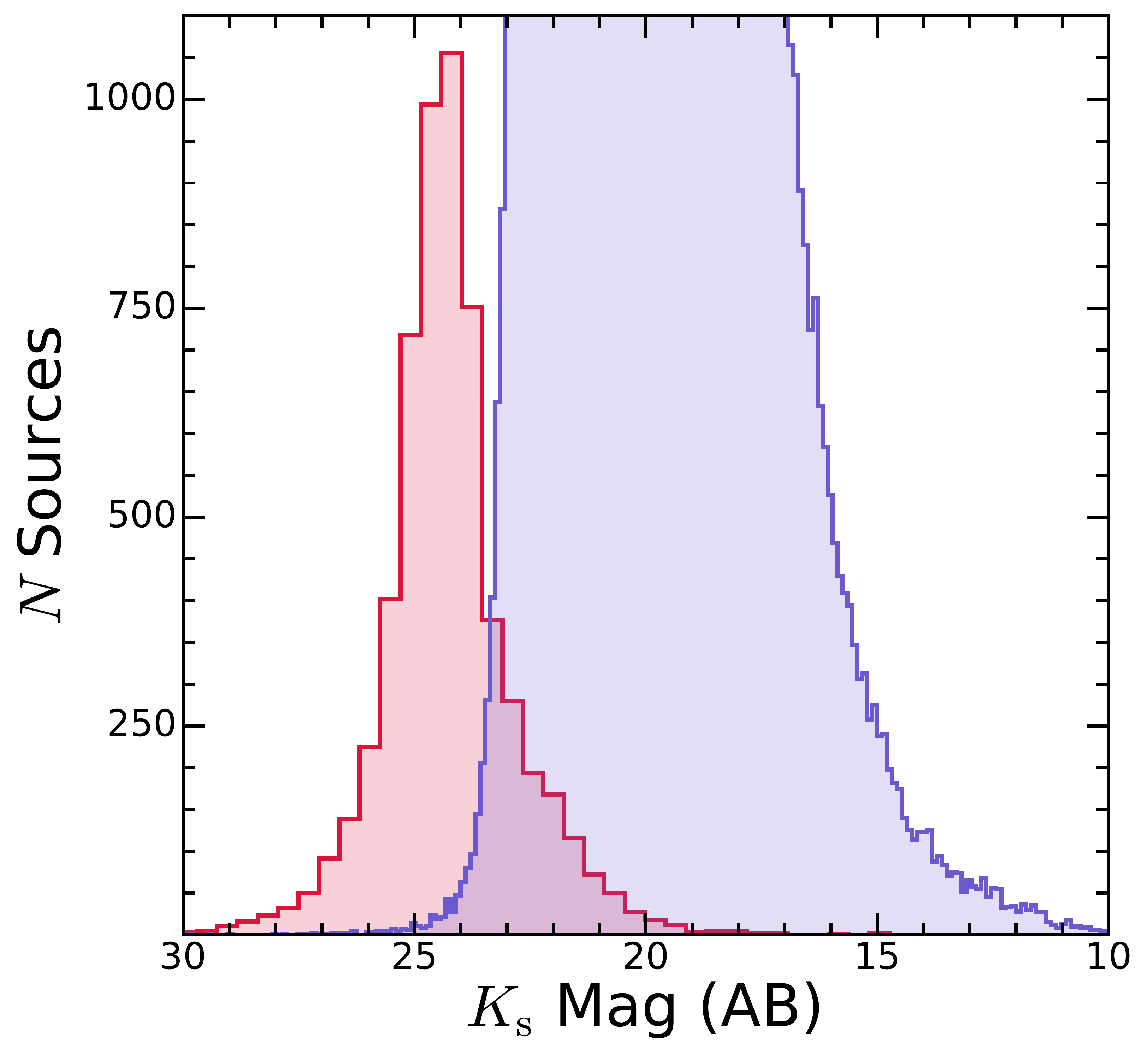}
\caption{{\bf Left:} Comparison of the number distribution of $K_{\mathrm{s}}$-band source magnitudes from the original VIDEO catalog (purple) and our forced photometry measurements based on the IRAC-selected input catalog (red).  The VIDEO-selected histogram represents Petrosian source magnitudes.  We emphasize that sources with measurements based on our IRAC-selected forced photometry were not detected in the original VIDEO catalog.  {\bf Right:} A zoomed-in view of the left panel that highlights the distribution of $K_{\mathrm{s}}$-band source magnitudes from our new IRAC-selected forced photometry.  %The vertical dashed line marks the original VIDEO $K_{\mathrm{s}}$-band 5$\sigma$ detection threshold from Table~\ref{tab:bands}.
\\}
\label{fig:IRAC_selected_dist}
\end{figure*}
%%%%%%%%%%%%%%%%%%%%%%%%%%%%%%%%%%%%%%%%%%%%%%%%

Since the IRAC bands themselves were used as priors, no de-blending was possible.  The primary benefit of performing forced photometry on the IRAC-selected catalog is the ability to identify faint VIDEO and CFHTLS-D1 counterparts to extremely red sources only detected previously in the IRAC bands.  Of the 8,441 sources in the IRAC-selected input catalog\footnote{We refer readers to Section~\ref{sec:input_cats} for details on the construction of the IRAC-selected input catalog.}, photometric measurements at $K_{\mathrm{s}}$-band were possible using {\it The Tractor} for $\approx69$\% of the sample.  We emphasize that this population of new $K_{\mathrm{s}}$-band detections represents intrinsically faint sources that fall below the detection threshold in the VIDEO single-band catalogs, but can be successfully measured with our forced photometry approach.  

Figure~\ref{fig:IRAC_selected_dist} compares the number distribution of $K_{\mathrm{s}}$-band source magnitudes from the original VIDEO-selected input catalog with measurements from our new IRAC-selected forced photometric catalog.  The distribution of new $K_{\mathrm{s}}$-band source magnitudes clearly demonstrates that our forced photometry detects a population of extremely faint, red objects that fall below the single-band detection threshold in the original VIDEO photometry.  Such intrinsically faint sources at or slightly below the original image detection threshold can only be detected with statistical techniques such as forced photometry that incorporate prior information about the source position from a detection at another band.

%%%%%%%%%%%%%%%%%%%%%%%%%%%%%%%%%%%%%%%%%%%%%%%%
\subsection{Depth}
\begin{figure*}[t!]
\centering
\includegraphics[clip=true, trim=0.5cm 0cm 0cm 0cm, width=17.5cm]{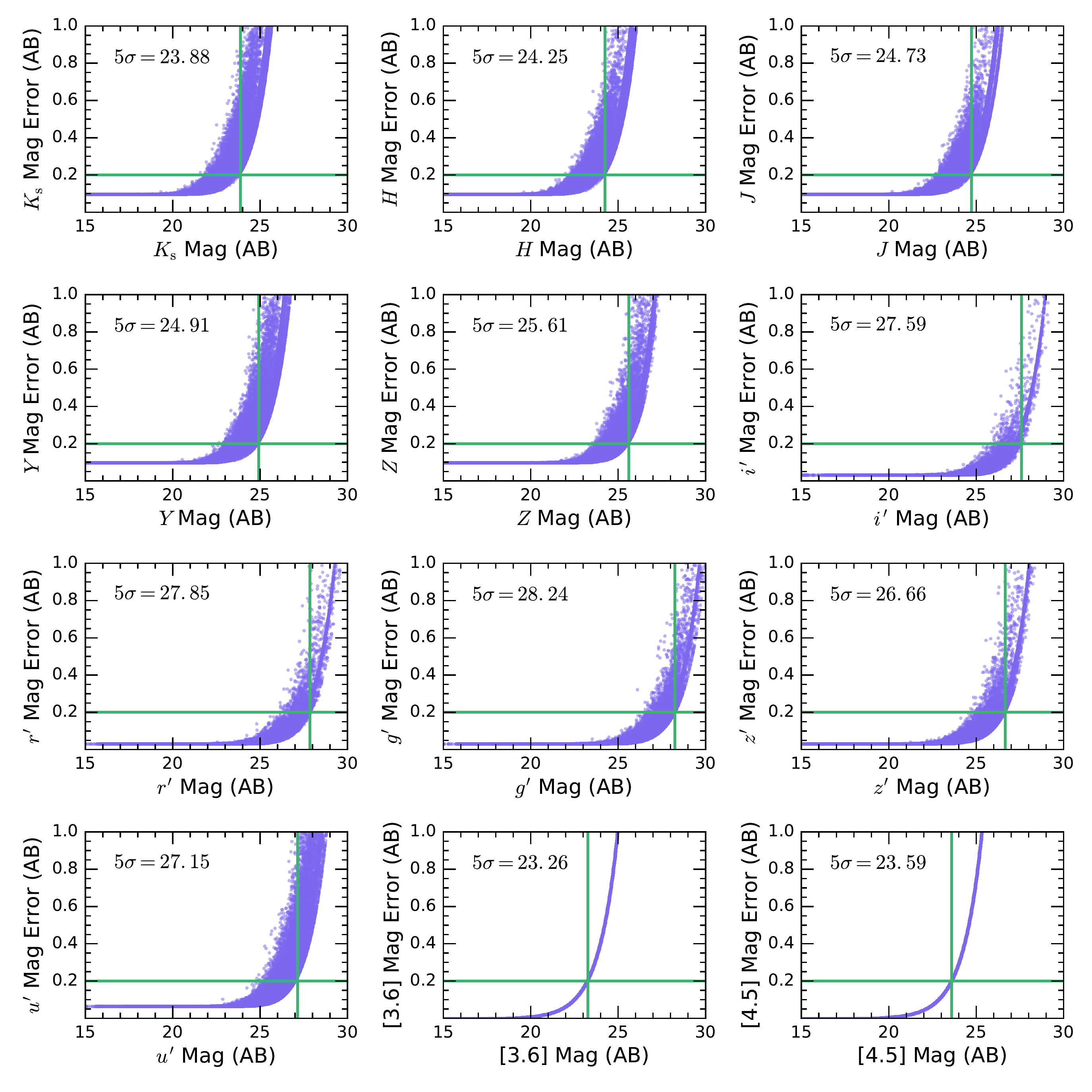}
\centering
\caption{Source magnitude versus magnitude error for the VIDEO-selected multi-band catalog produced using {\it The Tractor} to perform forced photometry.  The 5$\sigma$ magnitude limit corresponds to a magnitude error of 0.2 (horizontal green line).  For each band, we identify the faintest source magnitude at the intersection with the 5$\sigma$ limit (vertical green line).  We provide the value of the 5$\sigma$ detection threshold for each band in the upper left corner of each plot.\\  
}
\label{fig:mag_mag_err_video_sel}
\end{figure*}

\begin{figure*}[t!]
\centering
\includegraphics[clip=true, trim=0.5cm 0cm 0cm 0cm, width=17.5cm]{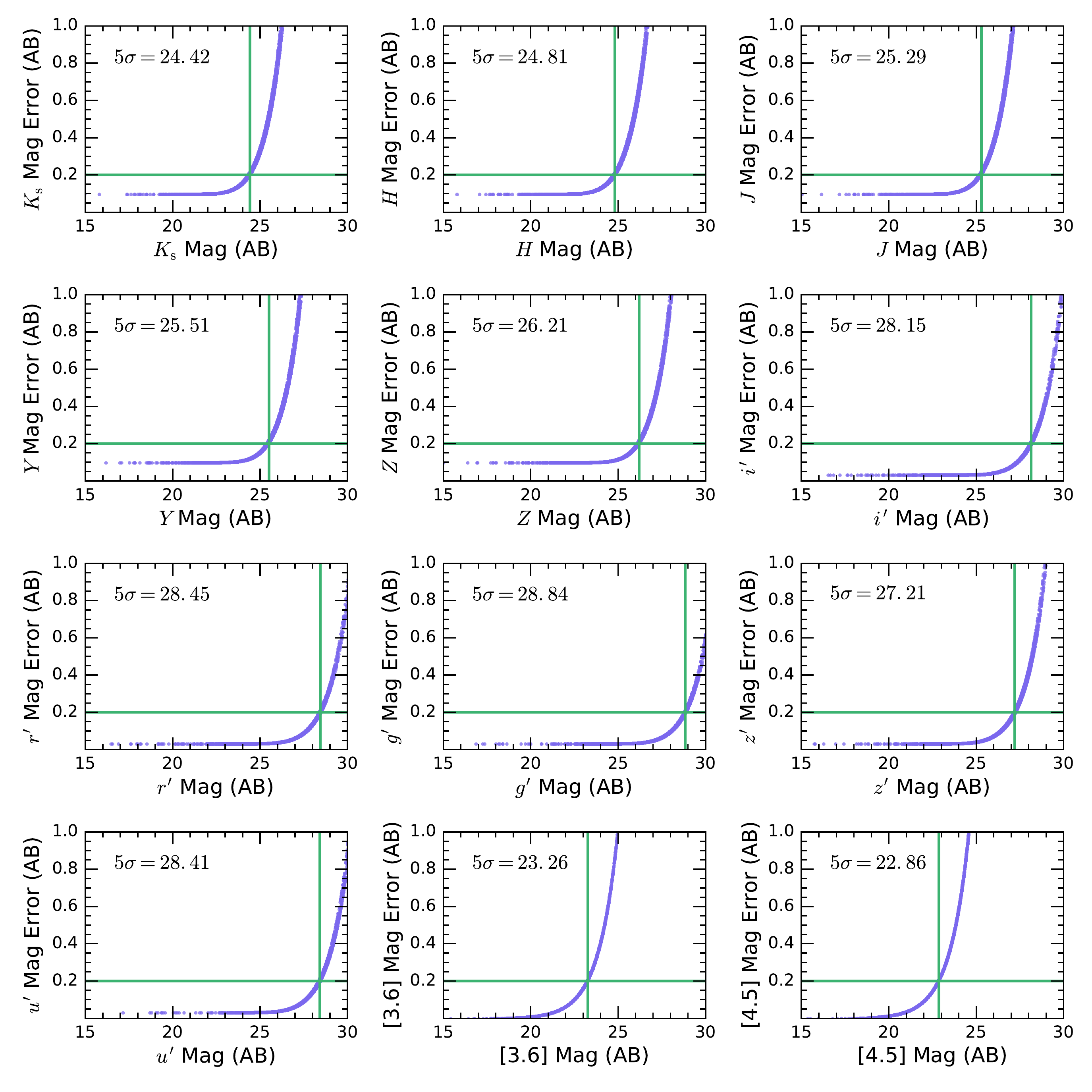}
\centering
\caption{Source magnitude versus magnitude error for the IRAC-selected multi-band catalog produced using {\it The Tractor} to perform forced photometry.  The 5$\sigma$ magnitude limit corresponds to a magnitude error of 0.2 (horizontal green line).  For each band, we identify the faintest source magnitude at the intersection with the 5$\sigma$ limit (vertical green line).  We provide the value of the 5$\sigma$ detection threshold for each band in the upper left corner of each plot.\\
}
\label{fig:mag_mag_err_irac_sel}
\end{figure*}

In Figures~\ref{fig:mag_mag_err_video_sel} and \ref{fig:mag_mag_err_irac_sel} we show the magnitude error\footnote{For a detailed discussion of the calculation of flux and magnitude errors, we refer readers to Appendix~\ref{sec:appendix_errors}.} as a function of magnitude for each band of the VIDEO- and IRAC-selected {\it Tractor} catalogs.  We use these figures to determine the 5$\sigma$ survey depth for each band by measuring the location in the distribution of magnitudes where the faintest sources reach a magnitude error of 0.2.  For the two IRAC bands at 3.6 and 4.5~$\mu$m, we find 5$\sigma$ limits of 23.26 and 23.59 in the VIDEO-selected output catalog, and 5$\sigma$ limits of 23.26 and 22.86 in the IRAC-selected catalog.  Values of the 5$\sigma$ depths for the remaining bands are provided in Figures~\ref{fig:mag_mag_err_video_sel} and \ref{fig:mag_mag_err_irac_sel}.  The 5$\sigma$ depths for our output forced photometry catalogs are generally comparable to the magnitude limits from the original catalogs shown in Table~\ref{tab:bands} or slightly deeper.

%%%%%%%%%%%%%%%%%%%%%%%%%%%%%%%%%%%%%%%%%%%%%%%%
%%%%%%%%%%%%%%%%%%%%%%%%%%%%%%%%%%%%%%%%%%%%%%%

%%%%%%%%%%%%%%%%%%%%%%%%%%%%%%%%%%%%%%%%%%%%%%%%
%%%%%%%%%%%%%%%%%%%%%%%%%%%%%%%%%%%%%%%%%%%%%%%

% CH1 - 14.0
% CH2 - 13.5

% Ks - N/A
% H - 14.0
% J - 14.5
% Y - 13.6
% Z - 13.8

% I - 16.3
% R - 15.9
% G - 15.9
% ZC - 15.1
% U - 15.7

%%%%%%%%%%%%%%%%%%%%%%%%%%%%%%%%%%%%%%%%%%%%%%%%
%%%%%%%%%%%%%%%%%%%%%%%%%%%%%%%%%%%%%%%%%%%%%%%%
%%%%%%%%%%%%%%%%%%%%%%%%%%%%%%%%%%%%%%%%%%%%%%%
\section{Discussion}
\label{sec:discuss}

%%%%%%%%%%%%%%%%%%%%%%%%%%%%%%%%%%%%%%%%%%%%%%%
\begin{figure*}[t!]
\includegraphics[clip=true, trim=0.1cm 0cm 0cm 0cm, width=8.5cm]{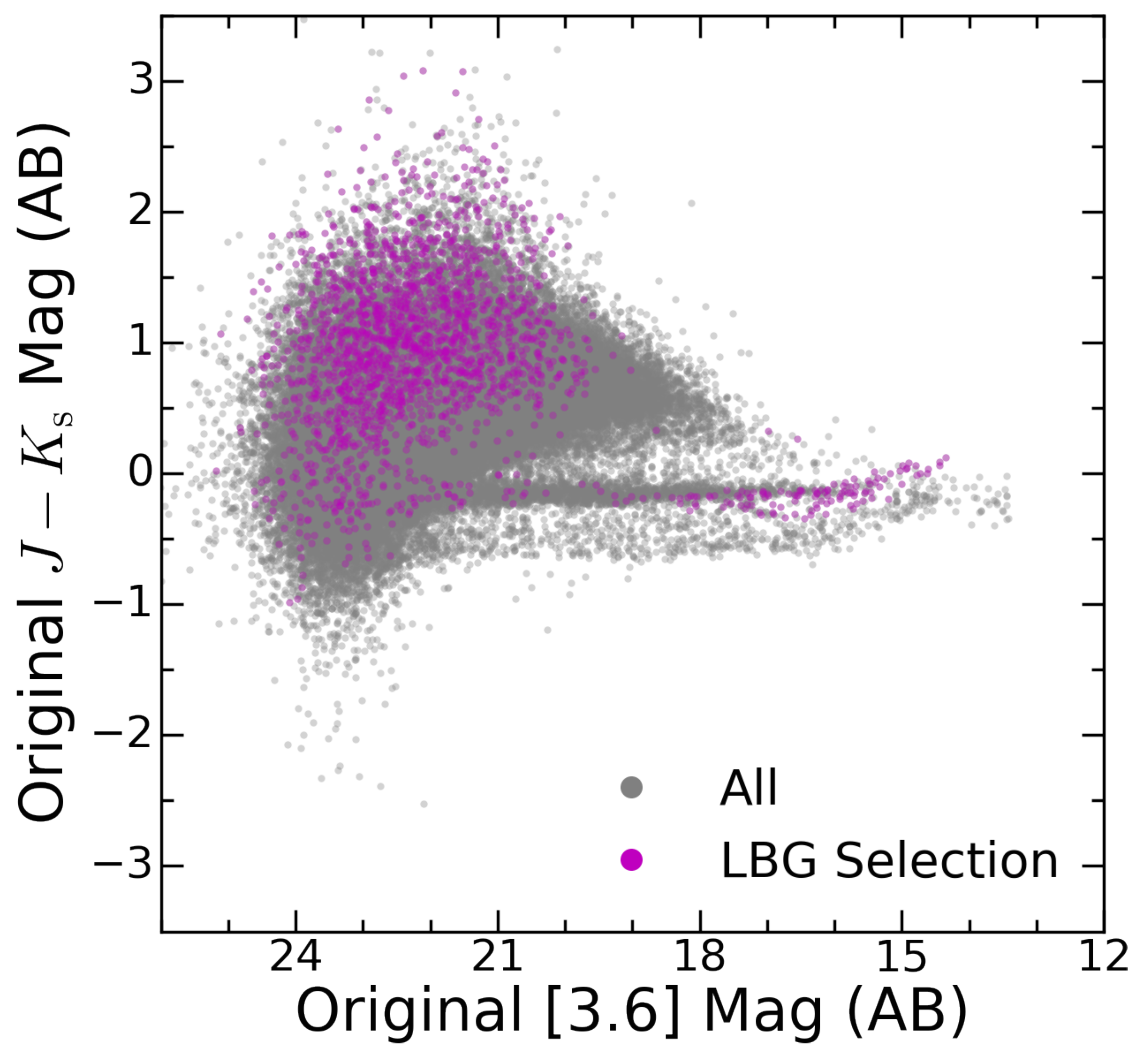}
\includegraphics[clip=true, trim=0.1cm 0cm 0cm 0cm, width=8.5cm]{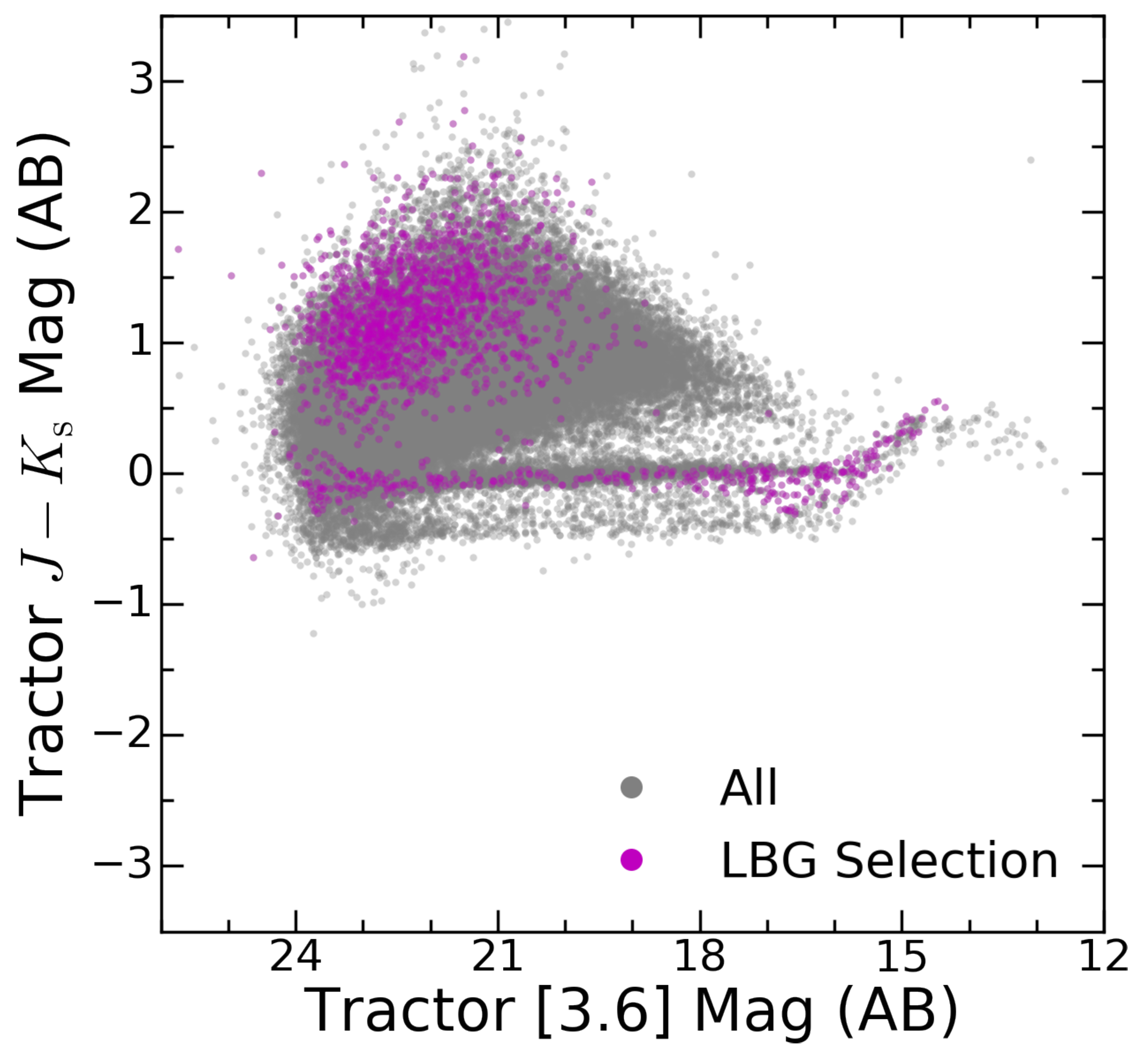}
\centering

\includegraphics[clip=true, trim=0.1cm 0cm 0cm 0cm, width=8.5cm]{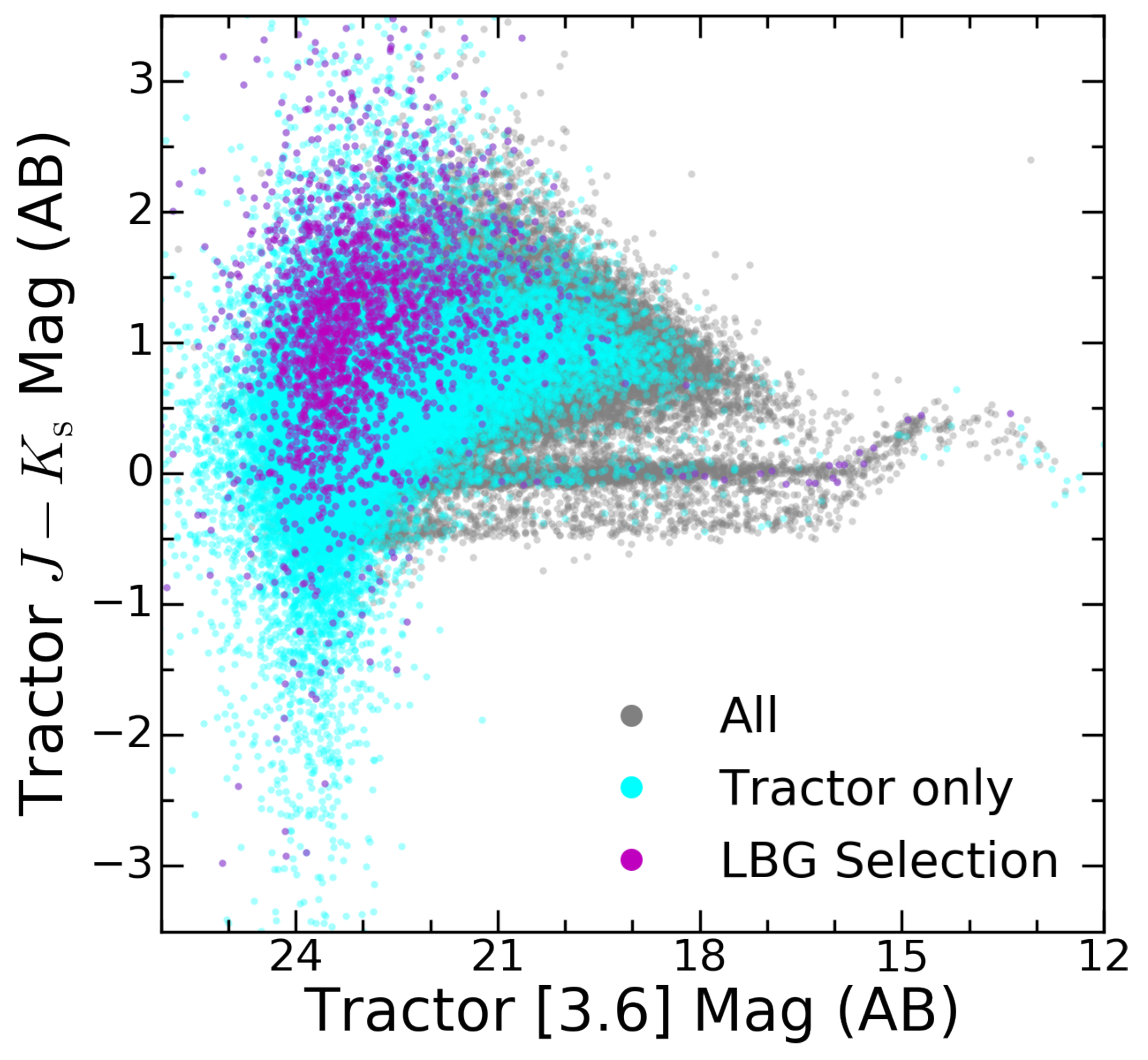}
\centering
\caption{{\bf Top left:} Comparison of the SERVS 3.6~$\mu$m magnitudes vs.\ the $J - K_{\mathrm{s}}$ VIDEO colors using the original photometry from the VIDEO-selected input source catalog (Section~\ref{sec:input_cats}).  Magnitudes are based on Petrosian, MAG\_AUTO, and 1.9$^{\prime \prime}$ apertures for VIDEO, CFHTLS-D1, and SERVS, respectively.  All sources with detections in the [3.6], $K_{\mathrm{s}}$, $J$, $u^{\prime}$, $g^{\prime}$, and $r^{\prime}$ bands (77,809 sources) are shown as gray symbols.  Sources that satisfy the $2.7 < z < 3.3$ LBG criteria of \citet{steidel+02} are highlighted in magenta.  The population of candidate LBG sources that lie on the stellar locus are likely low-redshift interlopers, such as Galactic halo main-sequence stars (e.g., K subdwarfs; \citealt{steidel+03}).
{\bf Top right:} Same as the top left panel, except here all magnitudes are based on our new forced photometry.  
{\bf Bottom:} Same as the middle panel, except the additional 30,198 sources that only have measurements in our new forced photometry catalog (i.e., those that were upper limits in the original input catalog) are shown in cyan.}
\bigskip
\label{fig:color_plots_1}
\end{figure*}
%%%%%%%%%%%%%%%%%%%%%%%%%%%%%%%%%%%%%%%%%%%%%%%%

%%%%%%%%%%%%%%%%%%%%%%%%%%%%%%%%%%%%%%%%%%%%%%%%
\subsection{Colors}
In Figure~\ref{fig:color_plots_1}, we show the 3.6 $\mu$m vs.\ $J - K_{\mathrm{s}}$ colors for the original VIDEO-selected input catalog (top left panel) and our new forced photometry (top right panel).  The same sources are shown in the top left and right panels - the only difference is the photometric catalog used to compute the colors.  Compared to the top left panel of Figure~\ref{fig:color_plots_1}, the top right panel showing the photometry from {\it The Tractor} has less scatter in the distribution of Lyman break galaxy (LBG)-selected sources ($g^{\prime} - r^{\prime} < 1.2$ and $u^{\prime} - g^{\prime} > g^{\prime} - r^{\prime} + 1$; \citealt{steidel+02}) residing in the redshift range $2.7 < z < 3.3$.  Figure~\ref{fig:color_plots_1} also clearly indicates that the stellar locus at $J - K_{\mathrm{s}} \approx -0.2$ is substantially better defined when the source colors are computed using {\it The Tractor} photometry.  This qualitatively suggests that the colors, and therefore the underlying photometric measurements, are more robust in our forced photometry catalog compared to the original input catalog.  We provide a more quantitative assessment of the improved photometric accuracy of our forced photometry catalog in Section~\ref{sec:photoz}.

In the bottom panel of Figure~\ref{fig:color_plots_1}, we show the NIR colors based on our forced photometry as in the middle panel, but this time we also highlight sources that were not detected in the original catalog.  This population of sources that are only identified in our new VIDEO-selected forced photometric catalog has a large degree of scatter, but this is expected given the intrinsically faint nature of many of these objects.  We emphasize that some of these sources do in fact lie within the main locus of galaxy colors, but were simply too faint to be detected in the original VIDEO photometry.  Thus, for studies geared towards intrinsically faint and potentially rare source populations, our implementation of photometry with {\it The Tractor} offers improved sensitivity compared to traditional positional matching methods.

%%%%%%%%%%%%%%%%%%%%%%%%%%%%%%%%%%%%%%%%%%%%%%%%
\subsection{Photometric Redshifts}
\label{sec:photoz}

%%%%%%%%%%%%%%%%%%%%%%%%%%%%%%%%%%%%%%%%%%%%%%%%
\subsubsection{Distribution}
One of the primary motivations for improving the accuracy of the original multi-band photometry is to obtain more robust photometric redshifts.  To test whether we have accomplished this in our test field, we have derived photometric redshifts based on the 12 NIR and optical SERVS, VIDEO and CFHTLS-D1 data described in this study using {\it HyperZ} \citep{bolzonella+00}.  Our galaxy SED template set-up follows that of \citet{pforr+13}, which is based on stellar population models from \citet{maraston+05}.  A detailed description of our application of SED fitting and subsequent determination of photometric redshifts will be presented in Pforr et al.\ (in preparation).  

%%%%%%%%%%%%%%%%%%%%%%%%%%%%%%%%%%%%%%%%%%%%%%%
\begin{figure*}[t!]
\centering
\includegraphics[clip=true, trim=0cm 0cm 0cm 0cm, width=8.9cm]{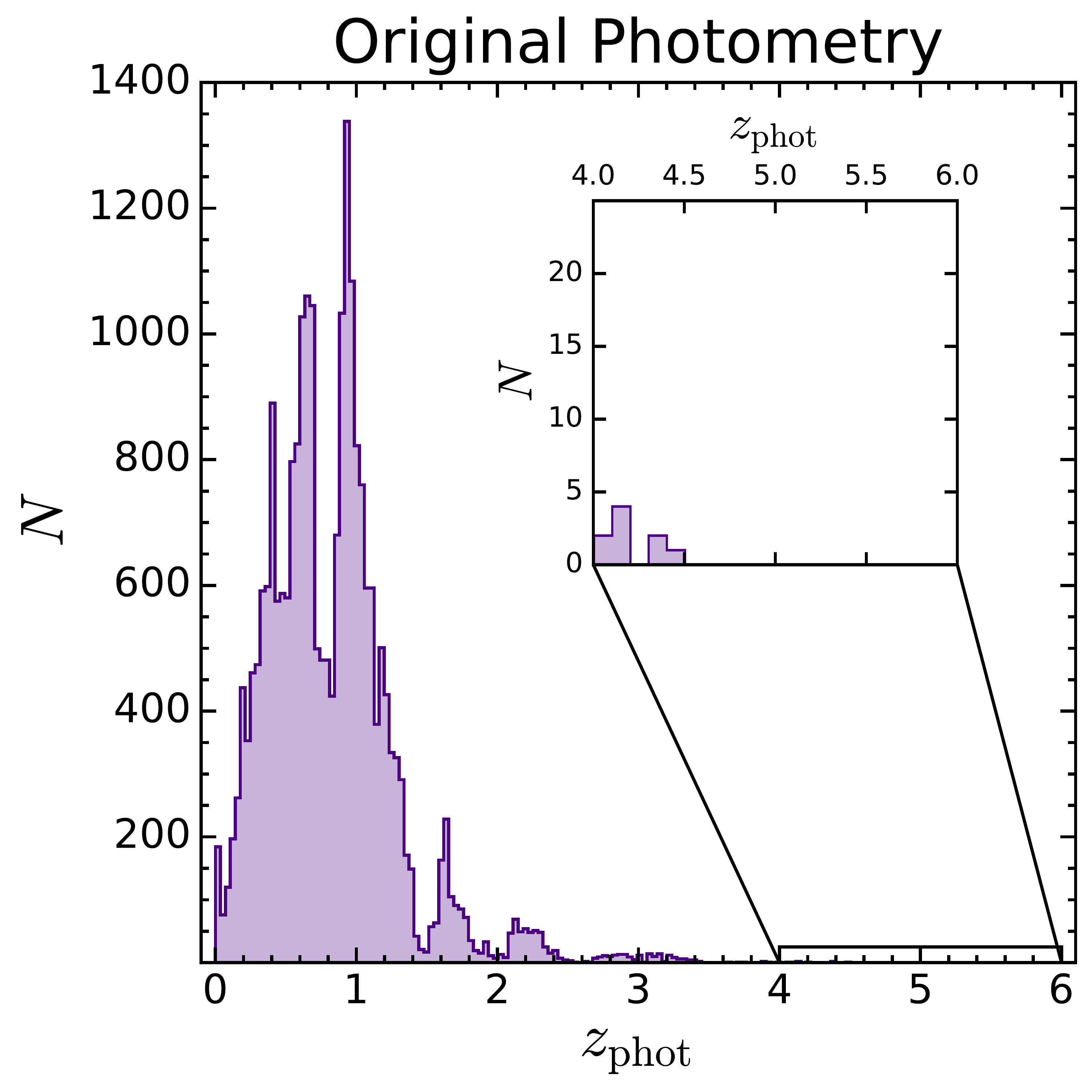}
\includegraphics[clip=true, trim=0cm 0cm 0cm 0cm, width=8.9cm]{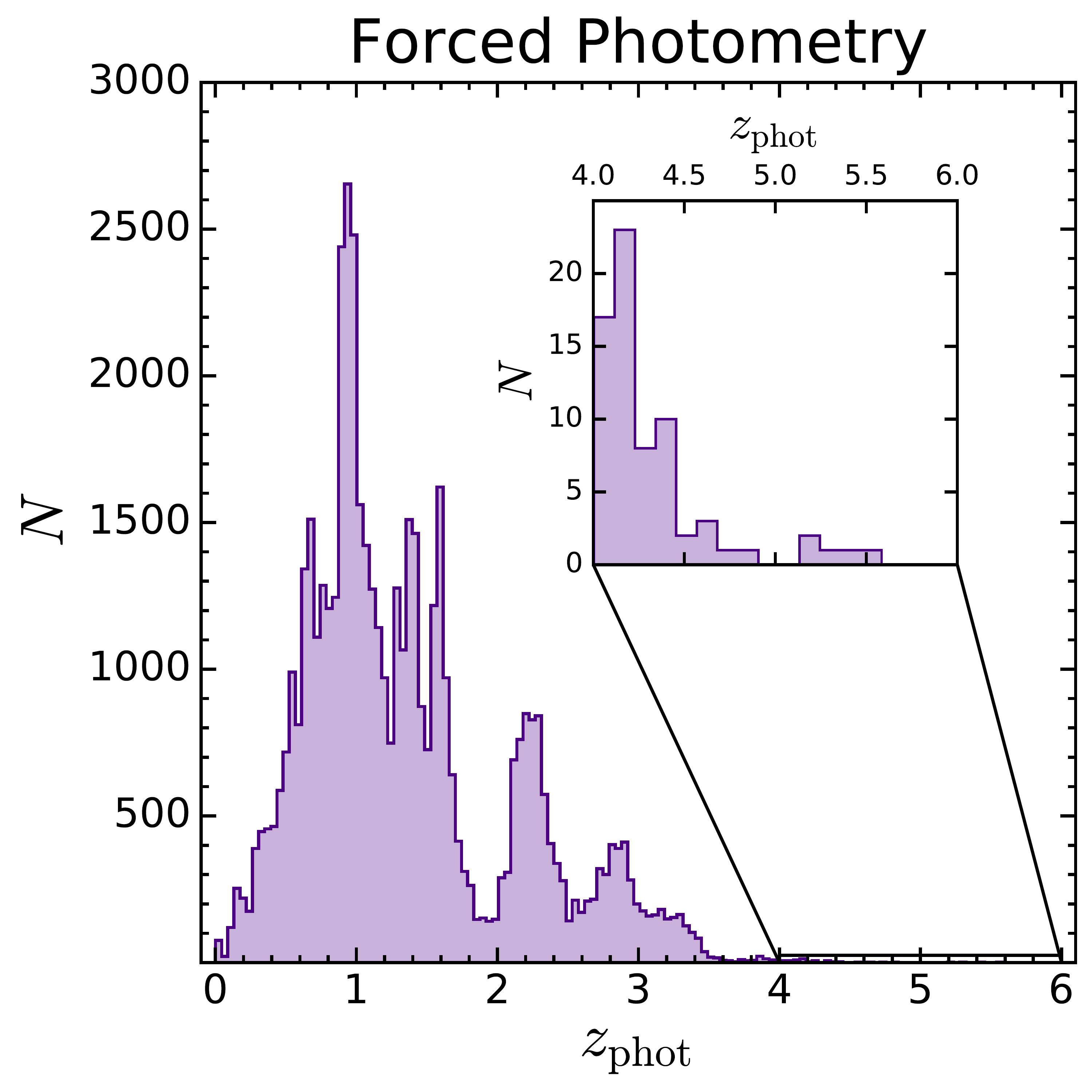}
\caption{{\bf Left:} Distribution of photometric redshifts, $z_{\mathrm{phot}}$, calculated using {\it HyperZ} based on the original, positional-matched source catalog in the square degree test region of XMM-LSS.  Details on the calculation of the photometric redshifts are provided in Pforr et al. (in preparation).  Only the 24,273 sources with measurements available in all 12 bands and accurate ($\chi^{2}_{\mathrm{red}} \lesssim 3.0$) photometric redshifts are included.  The inset axis shows the distribution of the 9 sources with high-redshifts in the range $4 < z_{\mathrm{phot}} < 6$.  
{\bf Right:} Same as the left panel, except here sources with accurate photometric measurements in all 12 bands from our new forced photometric catalog based on {\it The Tractor}.  Here, a total of 52,166 sources meet the criteria and are shown on the main axis.  The inset axis highlights the distribution of 70 sources with high redshifts. \\}
%\bigskip
\label{fig:phot_dist}
\end{figure*}
%%%%%%%%%%%%%%%%%%%%%%%%%%%%%%%%%%%%%%%%%%%%%%%

In Figure~\ref{fig:phot_dist}, we show the number distribution of photometric redshifts based on the original position-matched source catalogs and the new catalogs constructed using {\it The Tractor}.  For each catalog, only sources with accurate ($\chi^{2}_{\mathrm{red}} \leq 3.0$) photometric redshifts and measurements in all 12 bands are shown.  Both distributions are clearly dominated by lower-redshift sources, in harmony with the high proportion of sources modeled with resolved surface brightness profiles described in Section~\ref{sec:vid_sel_cat} that are typically associated with lower-redshift objects.  We note that the predominance of lower-redshift sources is not unexpected given that this redshift range covers the largest volume of our survey.  Due to our sensitivity limitations, we only detect the most luminous galaxies in the highest redshift bins.

%%%%%%%%%%%%%%%%%%%%%%%%%%%%%%%%%%%%%%%%%%%%%%%
\begin{figure*}[t!]
\centering
\includegraphics[clip=true, trim=0.5cm 0cm 26cm 0cm, width=8.8cm]{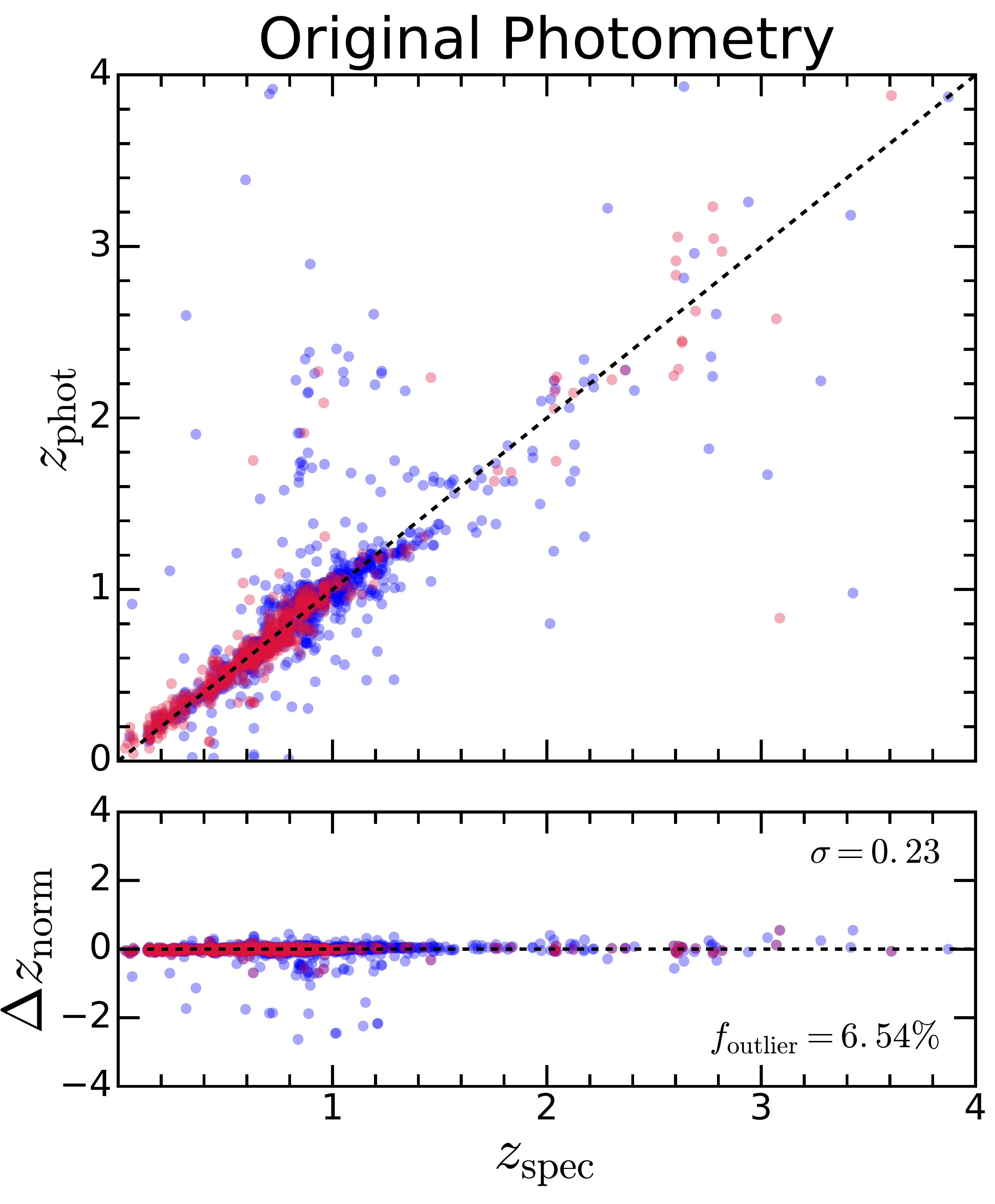}
\includegraphics[clip=true, trim=0cm 0cm 26cm 0cm, width=8.8cm]{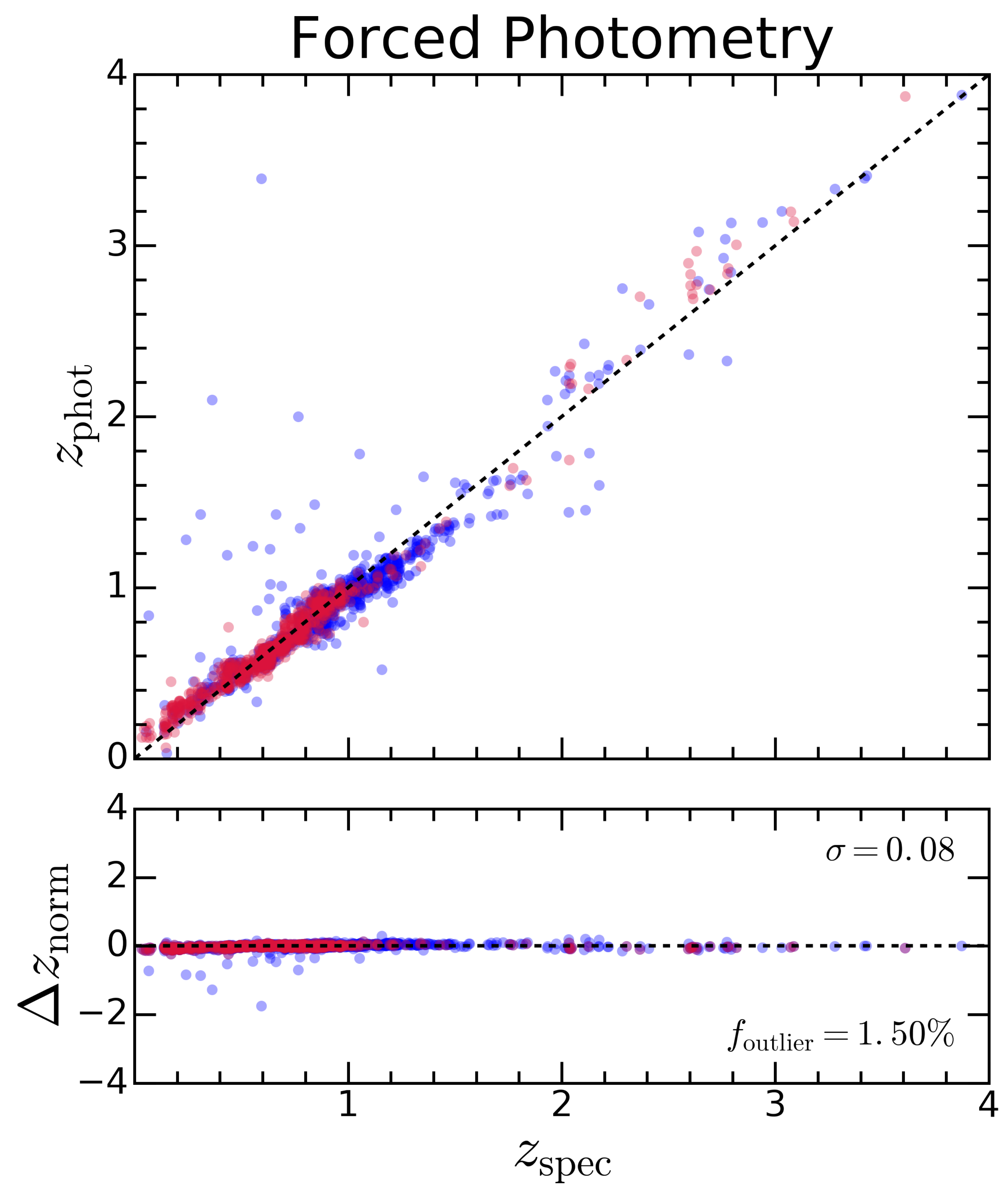}
\caption{{\bf Left:} Comparison of spectroscopic redshifts from the VIMOS VLT Deep Survey (VVDS; \citealt{le_fevre+13}) and the VIMOS Ultra Deep Survey (VUDS; \citealt{le_fevre+15}) with photometric redshifts determined from SERVS, VIDEO, and CFHTLS-D1 (Pforr et al., in preparation).  A total of 1,728 sources are shown.  Blue sources have spectroscopic redshifts with 95-100\% probability of being correct (flags 3 and 13) and red sources have spectroscopic redshifts that are highly certain with virtually 100\% probability of being correct (flags 4 and 14).  Only sources with accurate ($\chi^{2}_{\mathrm{red}} \leq 3.0$) photometric redshifts are included.  The lower panel shows the normalized residual, $\Delta z_{\mathrm{norm}}$ (Equation~\ref{Eq:res}), as a function of spectroscopic redshift.  The standard deviation, $\sigma$, and the fraction of outliers, $f_{\mathrm{outlier}}$ (Equation~\ref{Eq:outlier}), are also shown.
{\bf Right:} The exact same sources from the left panel are shown, except here the photometric redshifts are calculated using our new forced photometry.\\}
%\bigskip
\label{fig:spec_plots_1}
\end{figure*}
%%%%%%%%%%%%%%%%%%%%%%%%%%%%%%%%%%%%%%%%%%%%%%%

The comparison between the original and forced photometry-based photometric redshifts shown in this figure is striking.  When using the forced photometry source catalog, we obtain accurate photometric redshifts that incorporate all 12 bands into the SED fitting for over twice as many sources compared to the original position-matched photometry (52,166 vs.\ 24,273).  Furthermore, the number of high-redshift ($z > 4.0$) photometric redshifts sharply increases as well when the forced photometry is used.  Based on the original catalog, only 9 high-redshift sources are identified, though none of these are beyond $z = 5$.  In contrast, we find 70 candidate high-redshift sources when using our new forced photometry as input to {\it HyperZ}, 5 of which lie in the range $5 < z < 6$.  Thus, a clear advantage of using {\it The Tractor} is a substantial increase in the number of sources with robust photometric redshifts and improved sensitivity to faint, potentially high-redshift sources.

%%%%%%%%%%%%%%%%%%%%%%%%%%%%%%%%%%%%%%%%%%%%%%%%
\subsubsection{Spectroscopic Redshift Comparison}
Photometric redshifts from {\it HyperZ} based on the original, multi-band, position-matched catalogs and our new forced photometry are compared to high-quality spectroscopic redshifts\footnote{The VVDS and VUDS magnitude-limited redshift surveys are based on multi-slit spectroscopy over the wavelength range $3600 \lesssim \lambda \lesssim 9350$~$\AA$ and include galaxies up to redshift $z \sim 6.7$.} from the VIMOS VLT Deep Survey (VVDS; \citealt{le_fevre+13}) and the VIMOS Ultra-Deep Survey (VUDS; \citealt{le_fevre+15}) in Figure~\ref{fig:spec_plots_1}.  
The top left and right panels of this figure show photometric redshifts from the original VIDEO-selected input catalog and our new forced photometry, respectively.  As expected given the known prevalence of VIDEO sources that are blended in the SERVS photometry, Figure~\ref{fig:spec_plots_1} illustrates that the photometric and spectroscopic redshifts are much more tightly correlated when the photometric redshifts are determined using forced photometry.  
Figure~\ref{fig:spec_plots_1} also shows the standard deviation ($\sigma$) of the normalized residuals between the spectroscopic and photometric redshifts ($\Delta z_{\mathrm{norm}}$) and the outlier fraction ($f_{\mathrm{outlier}}$) for each photometric catalog.  These quantities are defined in Equations~\ref{Eq:res} and \ref{Eq:outlier} below:

%%%%%%%%%%%%%%%%%%%%%%%%%%%%%%%%%%%%%%%%%%%%%%%%
\begin{equation}
\label{Eq:res}
\Delta z_{\mathrm{norm}} = \frac{z_{\mathrm{spec}} - z_{\mathrm{phot}}}{1+z_{\mathrm{spec}}}\\
\end{equation}
%%%%%%%%%%%%%%%%%%%%%%%%%%%%%%%%%%%%%%%%%%%%%%%%

%%%%%%%%%%%%%%%%%%%%%%%%%%%%%%%%%%%%%%%%%%%%%%%%
\begin{equation}
\label{Eq:outlier}
f_{\mathrm{outlier}} = \left | \Delta z_{\mathrm{norm}} \right | > 0.15.
\end{equation}
%%%%%%%%%%%%%%%%%%%%%%%%%%%%%%%%%%%%%%%%%%%%%%%%

\medskip
For the 1,728 sources with accurate photometric redshifts ($\chi^{2}_{\mathrm{red}} \leq 3.0$), the original and forced photometry catalogs have $\sigma = 0.23$ and $\sigma = 0.08$, respectively.  This reduction in scatter for the forced photometry is consistent with the improvement in the $f_{\mathrm{outlier}}$ value, which is 6.54\% in the original photometry and 1.50\% in our new photometric catalog based on {\it The Tractor}.  To quantify the reduction in $\Delta z_{\mathrm{norm}}$ for the forced photometric catalog, we perform a two-sample Kolmogorov-Smirnov test \citep{feigelson+12} on $\Delta z_{\mathrm{norm}}$ from the original and {\it Tractor} catalogs.  This test yields a probability of $p = 5.6 \times 10^{-5}$ that the two samples are drawn from the same parent distribution, verifying that the reduction in the scatter for the normalized redshift residuals using forced photometry is indeed statistically significant.  This remarkable improvement is largely driven by the fact that {\it The Tractor} photometry provides photometric measurements for a larger number of bands included in our study compared to the original photometry based on the position-matched catalogs.  

%The photometric redshifts based on our new forced photometry are comparable in quality to those of previous studies utilizing many more bands.  For example, using a total of 30 bands ranging from the ultraviolet to the mid-infrared in the COSMOS 2 deg$^2$ field, \citet{ilbert+09} derived photometric redshifts with an outlier fraction of about 0.7\% for bright sources with $i^{+}$-band magnitudes $< 22.5$.  Our test field consists of considerably fainter sources with a median $i^{\prime}$-band magnitude of 23.0.  Yet, in spite of the fainter nature of our sources and the availability of at most 12 bands, our outlier fraction is only about a factor of two worse for the forced photometry-based photometric redshifts.  We conclude from our analysis of the photometric redshift accuracy of our sample that forced photometry with {\it The Tractor} provides substantially more accurate photometric redshifts compared to traditional multi-band, position-matching.

%%%%%%%%%%%%%%%%%%%%%%%%%%%%%%%%%%%%%%%%%%%%%%%%
%%%%%%%%%%%%%%%% FUTURE PROSPECTS %%%%%%%%%%%%%%%%%%%%
%%%%%%%%%%%%%%%%%%%%%%%%%%%%%%%%%%%%%%%%%%%%%%%%
\subsection{Future Science Applications}
\label{sec:science_apps}
The multi-band forced photometry of the VIDEO-selected input catalog provides a number of improvements over the original position-matched catalog in key areas including source matching accuracy, IRAC source de-blending, and sensitivity to faint sources below the single-band detection threshold in a given survey.  We have demonstrated that these improvements to the photometry lead to more accurate photometric redshifts, and in the future we plan to use our new forced photometric catalog to accurately measure galaxy masses to study stellar mass assembly out to $z \sim 5$.  We will also identify quasar candidates over a wide range of redshifts based on their NIR/optical colors (e.g., following an analysis similar to \citealt{richards+15}), taking advantage of our accurate photometry to study the demographics of obscured/unobscured quasars in different cosmic epochs.  This will allow us to assess the importance of AGN feedback and how it has evolved over the last 12 billion years. 

The forced photometry of the IRAC-selected input catalog, which contains sources with IRAC detections in the original SERVS photometry but no counterparts in any of the original VIDEO source catalogs, showed substantial improvement in the number of source detections in the VIDEO and CFHTLS-D1 bands.  At $K_{\mathrm{s}}$-band alone, the source detection fraction increased dramatically from 0\% in the original catalogs to 69\% after performing forced photometry with {\it The Tractor}.  This has important implications for the study of extremely red objects (EROs) that are detected in one or more of the SERVS bands but are not detected in any of the original VIDEO photometry.  EROs are believed to be extremely dust-enshrouded, high-redshift galaxies with high star formation rates, and represent an evolutionary stage of rapid assembly (e.g., \citealt{yan+04, wang+12}).  Despite the relevance of these objects to our understanding of galaxy formation and growth, large samples of EROs are currently lacking.  A future analysis of the SEDs and photometric redshifts of EROs identified in SERVS and analyzed with our implementation of {\it The Tractor} will provide much needed information on the properties and demographics of these objects, and address important galaxy evolution questions such as the fraction of obscured star formation missed by optical surveys.

We plan to expand our forced photometry implementation to the entire XMM-LSS field as well as the remaining four SERVS fields.  Given the availability of comparatively deep ground-based NIR and optical data, this will lead to accurate photometric redshift measurements over a large sky footprint, allowing us to maximize the scientific return of the SERVS project.  In the NIR, new data from the VISTA Extragalactic Infrared Legacy Survey (VEILS; \citealt{honig+17}) will provide additional deep data in the $J$ and $K_{\mathrm{s}}$ bands. 

%Deep optical imaging have recently been made publicly available by the Hyper Suprime-Cam Subaru Strategic Program Data Release 1 \citep{aihara+17}, which will allow us to perform forced photometry on the full XMM-LSS field along with the EN1 field later in 2017.  The first data release of the Panoramic Survey Telescope and Rapid Response System (Pan-STARRS) catalog of optical imaging over 5 bands covering 3$\pi$ steradian was recently made publicly available as well \citep{flewelling+16}.  However, to obtain comparable photometric redshift accuracy to that presented in this work for one square degree of the XMM-LSS field, we will require optical imaging of comparable depth to that of CFHTLS-D1 from the Pan-STARRS Medium Deep Survey \citep{huber+15}.  This survey covers four of the five SERVS fields (XMM-LSS, EN1, Lockman Hole, and CDFS), and is expected to be released later in the year.  For ES1, we must await deep optical imaging from the full-depth data release of the Dark Energy Survey (DES; \citealt{abbott+16}), which is expected to be made publicly available in 2020.

Deep optical imaging have recently been made publicly available by the Hyper Suprime-Cam Subaru Strategic Program Data Release 1 \citep{aihara+17}, which will allow us to perform forced photometry on the full XMM-LSS field along with the EN1 field later in 2017.  The first data release of the Panoramic Survey Telescope and Rapid Response System (Pan-STARRS) catalog of optical imaging over 5 bands covering 3$\pi$ steradian was recently made publicly available as well \citep{flewelling+16}.  However, to obtain comparable photometric redshift accuracy to that presented in this work for one square degree of the XMM-LSS field, we will require optical imaging of comparable depth to that of CFHTLS-D1 from the Pan-STARRS Medium Deep Survey \citep{huber+15}.  This survey covers four of the five SERVS fields (XMM-LSS, EN1, Lockman Hole, and CDFS), and is expected to be released later in the year.  Deep optical imaging over the full ES1 and CDFS fields has recently been made available as part of the VST Optical Imaging of the CDFS and ES1 Fields (VOICE; \citealt{vaccari+17}), and additional optical imaging of these fields (along with the XMM-LSS field) from the full-depth data release of the Dark Energy Survey (DES; \citealt{abbott+16}) is expected to be made publicly available in 2020.

We will also perform forced photometry on images from the {\it Spitzer} DEEPDRILL survey (P.I. Mark Lacy), which will provide post-cryogenic IRAC imaging to $\mu$Jy depth of the four predefined Deep Drilling Fields (DDFs) for the Large Synoptic Survey Telescope distributed over an area of 38.4 deg$^2$ (1 Gpc$^3$ at $z > 2$).  Science highlights of the DEEPDRILL survey include the detection of all the $>10^{11}$~M$_{\odot}$ galaxies out to $z \sim 6$ and the identification of $\sim40$ protoclusters at $z > 2$, which will provide numerous targets of interest for follow-up with {\it JWST}.  As is the case for SERVS, the legacy value of DEEPDRILL directly hinges upon the the availability of accurate multi-band photometry.  Thus, our application of forced photometry with {\it The Tractor} presented here will serve as an essential tool for ensuring the scientific success of both SERVS and DEEPDRILL, and will provide many important new insights into the physics of galaxy formation and evolution.

%%%%%%%%%%%%%%%%%%%%%%%%%%%%%%%%%%%%%%%%%%%%%%%
%%%%%%%%%%%%%%%%%%%%%%%%%%%%%%%%%%%%%%%%%%%%%%%
%%%%%%%%%%%%%%%%%%%%%%%%%%%%%%%%%%%%%%%%%%%%%%%
%%%%%%%%%%%%%%%%%%%%%%%%%%%%%%%%%%%%%%%%%%%%%%%
%%%%%%%%%%%%%%%%%% SUMMARY %%%%%%%%%%%%%%%%%%%%%%%
%%%%%%%%%%%%%%%%%%%%%%%%%%%%%%%%%%%%%%%%%%%%%%%
%%%%%%%%%%%%%%%%%%%%%%%%%%%%%%%%%%%%%%%%%%%%%%%
%%%%%%%%%%%%%%%%%%%%%%%%%%%%%%%%%%%%%%%%%%%%%%%
\section{Summary}
\label{sec:summary}
We have provided a description of our parallelized implementation of {\it The Tractor} to perform forced photometry on 12 NIR and optical bands from SERVS, VIDEO, and CFHTLS-D1 over a square degree of the XMM-LSS field.  The VIDEO- and IRAC-selected input catalogs -- which have 117,281 and 8,441 sources, respectively -- are used to define the fiducial source positions that establish the location at which a given source is modeled in each band.  For the VIDEO-selected input catalog, we found that use of {\it The Tractor} lead to the following key advantages compared to position-matched multi-band photometry:

\begin{enumerate}

\item By modeling the surface brightness profile of each source in a fiducial, high-resolution VIDEO band and performing forced photometry with {\it The Tractor}, we were able to de-blend these objects in the SERVS IRAC images and more accurately measure their photometric properties.  This naturally lead to more accurate source cross-identification, as evidenced by the improved definition of the stellar locus for {\it The Tractor} photometry shown in the color-color plot comparison in Figure~\ref{fig:color_plots_1}.  The importance of these improvements is highlighted by our estimated lower limit of 17\% for the number of sources that are clearly resolved in the VIDEO images, but blended in the lower-resolution SERVS data.  

\item Our application of multi-band forced photometry provided a higher fraction of source detections in each band.  This resulted in a factor of two increase in the number of sources with photometric redshift measurements with constraints in all of our optical/NIR bands (Figure~\ref{fig:phot_dist}).  As a direct consequence of this, we were able to identify a greater number of candidate high-redshift sources in our square degree test region.  While our new forced-photometry-based photometric redshifts identified 70 objects in the redshift range of $5 < z < 6$, the position-matched catalogs detected none.
%Thus, while photometric redshifts based on the position-matched, multi-band catalogs resulted in no candidate sources with redshifts in the range $5 < z < 6$, our new forced-photometry-based photometric redshifts identified 70 objects in that redshift range.

\item Based on comparisons between the photometric redshifts derived from the position-matched and forced photometry catalogs with spectroscopic redshift measurements from the literature, we found that {\it The Tractor} multi-band photometry lead to a statistically significant improvement in photometric redshift accuracy (Figure~\ref{fig:spec_plots_1}).  This will motivate follow-up analyses of various galaxy properties in our square-degree test region, such as stellar mass, for which significantly more accurate measurements can now be made.  
%Thus, our application of forced photometry with {\it The Tractor} will allow us to ultimately pursue exciting science applications, such as investigating the cosmic evolution and impact of various populations of quasars and AGNs, that would otherwise have been limited by the availability of accurate photometric redshifts.

\end{enumerate}

For the IRAC-selected input catalog, we found a dramatic improvement in the fraction of red objects with multi-band detections (from 0\% to 69\% at $K_{\mathrm{s}}$-band) in the VIDEO/CFHTLS-D1 data that were previously identified only in the SERVS imaging.  This opens up exciting new prospects for multi-wavelength analyses -- including photometric redshift estimates -- for large samples of extremely dust-enshrouded galaxies at high redshift.

In the future, we plan to apply our implementation of multi-band forced photometry to the full SERVS footprint as well as to new post-cryogenic {\it Spitzer} surveys such as DEEPDRILL that are currently in progress.  The clear improvement in photometric redshift accuracy that we have demonstrated will ultimately allow us to robustly address a number of key science topics in field of galaxy evolution, including stellar mass assembly, the fraction of obscured star formation, massive black hole growth and feedback, and the role of environment as galaxies grow and evolve.

%%%%%%%%%%%%%%%%%%%%%%%%%%%%%%%%%%%%%%%%%%%%%%
%%%%%%%%%%%%%%% ACKNOWLEDGMENTS %%%%%%%%%%%%%%%%%%%
%%%%%%%%%%%%%%%%%%%%%%%%%%%%%%%%%%%%%%%%%%%%%%
%\acknowledgments{{\it Acknowledgments:}
\section*{Acknowledgments}
We thank the referee for providing us with thoughtful comments that have significantly improved the quality of this work.  We also thank Scott Ransom for assisting us with the implementation of parallelization in our {\sc Python} driver script.  The National Radio Astronomy Observatory is a facility of the National 
Science Foundation operated under cooperative agreement by Associated Universities, Inc.  GW acknowledges financial support for this work from NSF grant AST-1517863 and from NASA through programs GO-13306, GO-13677, GO-13747 \& GO-13845/14327 from the Space Telescope Science Institute, which is operated by AURA, Inc., under NASA contract NAS 5-26555.  MV acknowledges support from the European Commission Research Executive Agency (FP7-SPACE-2013-1 GA 607254), the South African Department of Science and Technology (DST/CON 0134/2014) and the Italian Ministry for Foreign Affairs and International Cooperation (PGR GA ZA14GR02).

This work is based on observations made with the Spitzer Space Telescope, which is operated by the Jet Propulsion Laboratory, California Institute of Technology under a contract with NASA.  Support for this work was provided by the grant associated with {\it Spitzer} proposal 11086.  Our analysis includes observations obtained with the MegaPrime/MegaCam instrument, a joint project of CFHT and CEA/IRFU, at the Canada-France-Hawaii Telescope (CFHT) which is operated by the National Research Council (NRC) of Canada, the Institut National des Science de l'Univers of the Centre National de la Recherche Scientifique (CNRS) of France, and the University of Hawaii. This study is also based in part on data products produced at Terapix available at the Canadian Astronomy Data Centre as part of the Canada-France-Hawaii Telescope Legacy Survey, a collaborative project of NRC and CNRS.  We additionally utilized data from the VIMOS VLT Deep Survey, obtained from the VVDS database operated by Cesam, Laboratoire d'Astrophysique de Marseille, France.  

The authors have made use of {\sc Astropy}, a community-developed core {\sc Python} package for Astronomy \citep{astropy+13}.  We also used {\sc Montage}, which is funded by the National Science Foundation under Grant Number ACI-1440620, and was previously funded by the National Aeronautics and Space Administration's Earth Science Technology Office, Computation Technologies Project, under Cooperative Agreement Number NCC5-626 between NASA and the California Institute of Technology.  

\vspace{5mm}
\facilities{Spitzer (IRAC), VISTA, CFHT}

%\software{astropy \citep{2013A&A...558A..33A},  
%          Cloudy \citep{2013RMxAA..49..137F}, 
%          SExtractor \citep{1996A&AS..117..393B}
%          }

% List of Software: Tractor, Astropy, Montage, TOPCAT

%\software{{\it The Tractor} \citep{lang+16}, {\sc Astropy} \citep{astropy+13}, {\sc Montage} \citep{?}, TOPCAT \citep{?}
%}

%%%%%%%%%%%%%%%%%%%%%%%%%%%%%%%%%%%%%%%%%%%%%%%
%%%%%%%%%%%%%%%%%% BIBLIOGRAPHY %%%%%%%%%%%%%%%%%%%%
%%%%%%%%%%%%%%%%%%%%%%%%%%%%%%%%%%%%%%%%%%%%%%%
\bibliographystyle{apj}
\bibliography{XMM_sq_deg_tractor_v11}

%%%%%%%%%%%%%%%%%%%%%%%%%%%%%%%%%%%%%%%%%%%%%%%
%%%%%%%%%%%%%%%%%%%% APPENDIX %%%%%%%%%%%%%%%%%%%%%
%%%%%%%%%%%%%%%%%%%%%%%%%%%%%%%%%%%%%%%%%%%%%%%
\appendix

%%%%%%%%%%%%%%%%%%%%%%%%%%%%%%%%%%%%%%%%%%%%%%%
\section{Photometric Errors}
\label{sec:appendix_errors}

%%%%%%%%%%%%%%%%%%%%%%%%%%%%%%%%%%%%%%%%%%%%%%%
\subsection{SERVS}
Photometric errors for the SERVS bands are estimated as follows:

\begin{equation}
\label{eq:errors}
\sigma_{\mathrm{source}} = 3 \times \sqrt{  A \sigma_{\mathrm{rms}}^2 + F/G  },
\end{equation}
where the factor of 3 approximately accounts for correlated noise between pixels\footnote{The constant value of 3 in Equation~\ref{eq:errors} encompasses various effects related to the presence of correlated noise between pixels, including the fact that we have down-sampled the native pixel size by factor of two from $1\farcs2$ to $0\farcs6$ and performed drizzling to form the final SERVS IRAC mosaics.}, 
$A = \pi r^2$ is the circular source area in pixels, $\sigma_{\mathrm{rms}}$ is the background rms noise value per pixel, $F$ is the source brightness in native image units (MJy~sr$^{-1}$), and $G$ is the IRAC weighted gain.  For all SERVS sources in {\it The Tractor} catalog drawn from the IRAC-selected input catalog, as well as unresolved sources in the VIDEO-selected input catalog, the source radius, $r$, in pixels is determined by $r_{\mathrm{unresolved}}$:

\begin{equation}
\label{eq:fitr}
r_{\mathrm{unresolved}} = \sqrt{ \Sigma w_{i} \sigma_{i}^2 },
\end{equation}
where $w_{i}$ and $\sigma_{i}$ are the $i^{\mathrm{th}}$ Gaussian weight and standard deviation parameters that characterize the composite Gaussian PSF models of each band described in Table~\ref{tab:cal_params}.  For resolved sources drawn from the VIDEO-selected input catalog, the source radius, $r$, is instead measured by:

\begin{equation}
\label{eq:fitr_res}
r_{\mathrm{resolved}} = \sqrt{ (2 \times r_{\mathrm{source}})^2 + \, r_{\mathrm{unresolved}}^2},
\end{equation}
where $r_{\mathrm{source}}$ is based on the source radius definition given in Section~\ref{sec:surf_bright_prof} and the factor of two accounts for the fact that $r_{\mathrm{source}}$ traces the half-light radius rather than the full radius.

For $\sigma_{\mathrm{rms}}$, or the sky noise, we measure a single value for each image of each band (see Table~\ref{tab:cal_params}), and convert from MJy~sr$^{-1}$ to $\mu$Jy per pixel (1 MJy~sr$^{-1} \approx 8.46$~$\mu$Jy~pixel$^{-2}$) before inserting this term into Equation~\ref{eq:errors}.  The weighted gain, $G$, is defined\footnote{For further details on the derivation of the weighted gain, $G$, we refer readers to Section~5.2 of \citet{timlin+16}.} by the following:

\begin{equation}
\label{eq:G}
G = \frac{N \times g \times T}{K},
\end{equation}
where $N$ is the average number of coverages estimated from each {\it Spitzer} Astronomical Observation Request (AOR) (the XMM-LSS SERVS observations consisted of 12 AORs), $g$ is the detector gain (3.70 $e^{-1}$(DN)$^{-1}$ for the 3.6~$\mu$m band and 3.71 $e^{-1}$(DN)$^{-1}$ for the 4.5~$\mu$m band), $T$ is the exposure time per coverage (100 seconds for SERVS), and $K$ is the conversion factor from digital to physical units.  
%Equation~\ref{eq:errors} encapsulates the rms background noise added in quadrature with the Poisson noise.  
The flux uncertainty for each source is converted from MJy~sr$^{-1}$ to $\mu$Jy~pixel$^{-2}$ using the relation $F(\mu$Jy~pixel$^{-2}) = 8.46 \times F($MJy~sr$^{-1}$).   %Finally, we increase the uncertainty in the [3.6] and [4.5] bands by additional factors of 3 and 2.5, respectively, such that sources with magnitudes near the 5$\sigma$ detection threshold have flux uncertainties of about 20\% and magnitude uncertainties of 0.2 mag.

%%%%%%%%%%%%%%%%%%%%%%%%%%%%%%%%%%%%%%%%%%%%%%%
\subsection{VIDEO and CFHTLS-D1}
For the ground-based VIDEO and CFHTLS-D1 bands, we estimate the photometric errors according to the formula below:

\begin{equation}
\label{eq:errors2}
\sigma_{\mathrm{source}} = C \times \sqrt{  A \sigma_{\mathrm{rms}}^2 + (0.03 \times F)^2  },
\end{equation}
where $A$ and $\sigma_{\mathrm{rms}}^2$ are defined as in Equation~\ref{eq:errors}, $F$ is the source flux in $\mu$Jy, and the factor of 0.03 accounts for the expected photometric error due to systematic uncertainties in the underlying image calibration.  The constant $C$ in Equation~\ref{eq:errors2} accounts for correlated pixel noise, and has a value of 3 for VIDEO and 1 for CFHTLS-D1.

%%%%%%%%%%%%%%%%%%%%%%%%%%%%%%%%%%%%%%%%%%%%%%%
\subsection{Magnitude Uncertainties}
Magnitude uncertainties for all bands are calculated following the standard conversion from the flux uncertainties using the equation below:

\begin{equation}
\label{eq:mag_err}
\sigma_{\mathrm{mag}} = 1.089 \times \sigma_{\mathrm{source}}/f_{\mathrm{source}},
\end{equation}
where $f_{\mathrm{source}}$ is the source brightness, and both $\sigma_{\mathrm{source}}$ and $f_{\mathrm{source}}$ are measured in units of $\mu$Jy.

\begin{deluxetable}{ccccc}[t!]
\tablecaption{Table columns and their descriptions for the VIDEO- and IRAC-selected output {\it Tractor} catalogs \label{tab:cat_columns}}
\tablecolumns{5}
%\tablenum{3}
\tablewidth{10pt}
\tablehead{
\colhead{Name} & \colhead{Description} & \colhead{  }& \colhead{Name} & \colhead{Description}
}
\startdata
RA & J2000 position from original VIDEO catalog & & 3.6mag & SERVS [3.6] magnitude\\ 
Dec & J2000 position from original VIDEO catalog & & 4.5mag & SERVS [4.5] magnitude\\ 
Flux-Ks & VIDEO $Ks$-band flux & & e\_Ksmag & VIDEO $Ks$-band magnitude error\\ 
Flux-H & VIDEO $H$-band flux & &  e\_Hmag & VIDEO $H$-band magnitude error \\ 
Flux-J & VIDEO $J$-band flux & & e\_Jmag & VIDEO $J$-band magnitude error \\ 
Flux-Y & VIDEO $Y$-band flux & & e\_Ymag & VIDEO $Y$-band magnitude error \\ 
Flux-Z & VIDEO $Z$-band flux & & e\_Zmag & VIDEO $Z$-band magnitude error \\ 
Flux-$i^{\prime}$ & CFHTLS-D1 $i^{\prime}$-band flux & & e\_$i^{\prime}$mag & CFHTLS-D1 $i^{\prime}$-band magnitude error \\ 
Flux-$r^{\prime}$ & CFHTLS-D1 $r^{\prime}$-band flux & & e\_$r^{\prime}$mag & CFHTLS-D1 $r^{\prime}$-band magnitude error \\ 
Flux-$g^{\prime}$ & CFHTLS-D1 $g^{\prime}$-band flux & & e\_$g^{\prime}$mag & CFHTLS-D1 $g^{\prime}$-band magnitude error  \\ 
Flux-$z^{\prime}$C & CFHTLS-D1 $z^{\prime}$-band flux & & e\_$z^{\prime}$Cmag & CFHTLS-D1 $z^{\prime}$-band magnitude error \\ 
Flux-$u^{\prime}$ & CFHTLS-D1 $u^{\prime}$-band flux & & e\_$u^{\prime}$mag & CFHTLS-D1 $u^{\prime}$-band magnitude error \\ 
Flux-3.6 & SERVS [3.6] flux & & e\_3.6mag & SERVS [3.6] magnitude error \\ 
Flux-4.5 & SERVS [4.5] flux & & e\_4.5mag & SERVS [4.5] magnitude error \\ 
e\_Flux-Ks & VIDEO $Ks$-band flux error & & redchi-Ks & VIDEO $Ks$-band $\chi^{2}_{\mathrm{red}}$ \\ 
e\_Flux-H & VIDEO $H$-band flux error & & redchi-H & VIDEO $H$-band $\chi^{2}_{\mathrm{red}}$ \\ 
e\_Flux-J & VIDEO $J$-band flux error & & redchi-J & VIDEO $J$-band $\chi^{2}_{\mathrm{red}}$ \\ 
e\_Flux-Y & VIDEO $Y$-band flux error & & redchi-Y & VIDEO $Y$-band $\chi^{2}_{\mathrm{red}}$ \\ 
e\_Flux-Z & VIDEO $Z$-band flux error & & redchi-Z & VIDEO $Z$-band $\chi^{2}_{\mathrm{red}}$ \\ 
e\_Flux-$i^{\prime}$ & CFHTLS-D1 $i^{\prime}$-band flux error & & redchi-$i^{\prime}$ & CFHTLS-D1 $i^{\prime}$-band $\chi^{2}_{\mathrm{red}}$ \\ 
e\_Flux-$r^{\prime}$ & CFHTLS-D1 $r^{\prime}$-band flux error & & redchi-$r^{\prime}$ & CFHTLS-D1 $r^{\prime}$-band $\chi^{2}_{\mathrm{red}}$ \\ 
e\_Flux-$g^{\prime}$ & CFHTLS-D1 $g^{\prime}$-band flux error & & redchi-$g^{\prime}$ & CFHTLS-D1 $g^{\prime}$-band $\chi^{2}_{\mathrm{red}}$ \\ 
e\_Flux-$z^{\prime}$C & CFHTLS-D1 $z^{\prime}$-band flux error & & redchi-$z^{\prime}$C & CFHTLS-D1 $z^{\prime}$-band $\chi^{2}_{\mathrm{red}}$ \\ 
e\_Flux-$u^{\prime}$ & CFHTLS-D1 $u^{\prime}$-band flux error & & redchi- $u^{\prime}$ & CFHTLS-D1 $u^{\prime}$-band $\chi^{2}_{\mathrm{red}}$ \\ 
e\_Flux-3.6 & SERVS [3.6] flux error & & redchi-3.6 & SERVS [3.6] $\chi^{2}_{\mathrm{red}}$ \\ 
e\_Flux-4.5 & SERVS [4.5] flux error  & & redchi-4.5 & SERVS [4.5] $\chi^{2}_{\mathrm{red}}$ \\ 
Ksmag & VIDEO $Ks$-band magnitude  & & FiducialBand & VIDEO band used to define source model \\ 
 Hmag & VIDEO $H$-band magnitude & & SourceModel & Source surface brightness profile model\\
 Jmag & VIDEO $J$-band magnitude & & Sat-flag & Source saturation flag in each band\\ 
 Ymag & VIDEO $Y$-band magnitude \\ 
 Zmag & VIDEO $Z$-band magnitude \\ 
 $i^{\prime}$mag & CFHTLS-D1 $i^{\prime}$-band magnitude \\ 
 $r^{\prime}$mag & CFHTLS-D1 $r^{\prime}$-band magnitude \\ 
 $g^{\prime}$mag & CFHTLS-D1 $g^{\prime}$-band magnitude \\ 
$z^{\prime}$Cmag & CFHTLS-D1 $z^{\prime}$-band magnitude \\ 
 $u^{\prime}$mag & CFHTLS-D1 $u^{\prime}$-band magnitude \\ 
%Mag\_err\_Ks & \\ 
%Mag\_err\_Ks & \\ 
%Mag\_err\_Ks & \\ 
%Mag\_err\_Ks & \\ 
%Mag\_err\_Ks & \\ 
%Mag\_err\_Ks & \\ 
%Mag\_err\_Ks & \\ 
%Mag\_err\_Ks & \\ 
%Mag\_err\_Ks & \\ 
%Mag\_err\_Ks & \\ 
%Mag\_err\_Ks & \\ 
%Mag\_err\_Ks & \\ 
%Fiducial\_band & \\ 
%Model & \\ 
\enddata
\tablecomments{All fluxes are given in units of $\mu$Jy and all magnitudes are based on the AB system.  A description of the calculation of flux and magnitude uncertainties is given in Appendix~\ref{sec:appendix_errors}.  SourceModel values may be one of the following: Dev (deVaucouleurs profile), Exp (exponential profile), or PtSrc (point source model).
For the Sat-flag column, values are a single binary string containing a 0 (not saturated) or a 1 (saturated) for each band in the following order: $Ks$, $H$, $J$, $Y$, $Z$, $i^{\prime}$, $r^{\prime}$, $g^{\prime}$, $z^{\prime}$, $u^{\prime}$, $3.6~\mu$m, and $4.5~\mu$m.  The Sat-flag column is only included for the VIDEO-selected output source catalog since saturation is not an issue in the IRAC-selected catalog of intrinsically fainter sources.}
\end{deluxetable}

%\subsection{IRAC-Selected Catalog}
%{\bf \color{red} DESCRIPTION OF CATALOG OUTPUT TO GO HERE.}

%!%!%!%!%!%!%!%!%!%!%!%!%!%!%!%!%!%!%!%!%!%!%!%!%!%!%!%!%!%!%!%!%!%!%!%!%!%!%!%!%!%!%!%!%!%!%!%!%!%!%!%!%!%!
%!%!%!%!%!%!%!%!%!%!%!%!%!%!%!%!%!%!%!%!%!%!%!%!%!%!%!%!%!%!%!%!%!%!%!%!%!%!%!%!%!%!%!%!%!%!%!%!%!%!%!%!%!%!
%!%!%!%!%!%!%!%!%!%!%!%!%!%!%!%!%!%!%!%!%!%!%!%!%!%!%!%!%!%!%!%!%!%!%!%!%!%!%!%!%!%!%!%!%!%!%!%!%!%!%!%!%!%!
%\bsp	% typesetting comment
%\label{lastpage}
\end{document}